\documentstyle[preprint,eqsecnum,aps]{revtex}

\def\aleq{\vcenter{\vbox{\hbox{$\buildrel < \over \sim$}}}}
\def\ageq{\vcenter{\vbox{\hbox{$\buildrel > \over \sim$}}}}
\begin{document}
\draft
\preprint{\parbox[b]{5.0cm}{UNIGRAZ-UTP 15-04-96 WU B 96-13}}
\title{Exclusive Photoproduction \protect \\ of Large Momentum-Transfer K and
K$^*$ Mesons}
\author{P. Kroll\cite{kro} and M. Sch\"urmann\cite{schue}}
\address{Fachbereich Physik, Universit\"at Wuppertal, D-42097 Wuppertal,
Germany}
\author{K. Passek\cite{pa}}
\address{Rudjer Bo\v{s}kovi\'c Institute, P.O. Box 1016, HR-10000 Zagreb,
Croatia}
\author{W. Schweiger\cite{schw}}
\address{Institut f\"ur Theoretische Physik, Universit\"at Graz, A-8010 Graz,
Austria}
\date{April 15, 1996}
\maketitle
\begin{abstract}
The reactions $\gamma \text{p} \longrightarrow \text{K}^{+} \Lambda$ and
$\gamma
\text{p} \longrightarrow \text{K}^{\ast +} \Lambda$ are analyzed within
perturbative QCD,
allowing for diquarks as quasi-elementary constituents of baryons. The
diquark-model
parameters and  the quark-diquark distribution amplitudes of proton and Lambda
are taken
from previous investigations of electromagnetic baryon form factors and
Compton-scattering
off protons. Unpolarized differential cross sections and polarization
observables are
computed for different choices of the K and K$^{\ast}$ distribution amplitudes.
The
asymptotic form of the K distribution amplitude ($\propto x_1 x_2$) is found to
provide a satisfactory description of the K photoproduction data.
\end{abstract}

\narrowtext

\section{Introduction}
\label{sec:intro}
The reactions $\gamma  \text{p} \rightarrow K  \text{Y}$
(Y=$\Lambda$,$\Sigma$) belong
to the most elementary processes which allow to study strangeness production.
Stimulated by
the advent of a new generation of intermediate energy electron facilities, like
ELSA or
CEBAF, they recently received renewed interest. Traditional hadronic models
applied to the
analysis of K photoproduction mostly make use of Feynman-diagram techniques
\cite{Be89,AS90,WJC92}. The corresponding reaction mechanism is based on the
exchange of p,
$\Lambda$, $\Sigma$, K, and K$^{\ast}$, along with a varying number of
N$^{\ast}$ and
Y$^{\ast}$ resonances. Apart from some problems with SU(3) bounds on the
hadronic
coupling constants $g_{\text{K} \text{Y} \text{N}}$ \cite{MBH95}, such models
seem to work
properly for photon energies  up to
$p_{\text{lab}}^\gamma \aleq 1.4 - 2.2$~GeV. New data of
higher precision and completeness (which include also spin observables)
\cite{CEBAF95,ELSA95}
are expected to restrict still persisting uncertainties in the meson-baryon
couplings and
the resonance parameters.

A more fundamental treatment of photoproduction should, of course, rely on QCD,
the dynamics
of interacting quarks and gluons. A step in this direction are effective,
"QCD-inspired"
models which include already one or the other feature of QCD. A particular
example is the
chiral quark model which has been applied to K photoproduction very recently
\cite{Li95}. Its
elementary degrees of freedom are constituent quarks and the members of the
(lowest lying)
pseudoscalar meson octet (K, $\eta$, and $\pi$). The latter are considered as
Goldstone
bosons associated with the spontaneous breaking of chiral symmetry. The 
Goldstone bosons couple directly to the (confined) quarks . With less 
parameters than hadronic
approaches this model also provides reasonable results for
$p_{\text{lab}}^\gamma \aleq 2$~
GeV. Its restricted range of validity is caused by the use of non-relativistic
transition
operators and baryon wave functions.

Direct application of QCD is (till now) restricted to kinematical situations in
which
the scattering of the hadronic constituents and their hadronization takes place
on rather
different scales. In general this means large energies and momentum transfers
($p_{\perp}$). Well
beyond the resonance region ($p_{\text{lab}}^\gamma \gg 1$~GeV) exclusive
photoproduction cross
sections exhibit a characteristic angular dependence. At forward (small $t$)
and backward
(small $u$) angles the strong variation of the differential cross section is
adequately
reproduced by the exchange of meson and baryon Regge trajectories, respectively
\cite{Do72}.
Around $\theta_{\text{cm}} = 90^{\circ}$ (large $t$ and $u$) the cross section
flattens and shows (for
fixed angles) an energy dependence typical for a hard interaction between the
photon and the
constituents inside the proton. A constituent scattering model for high-energy,
large-$p_{\perp}$
elastic and quasielastic reactions has been proposed in Ref.~\cite{GBB73}. The
interaction mechanism
of this model, namely quark interchange, may be also thought of as one of the
simplest ways to
describe photoproduction of open strangeness. The resulting interchange
amplitude is just a
convolution over light-cone wave functions of the interacting quarks, which
have
to be parameterized in
an appropriate way.

A more subtle picture, often called the \lq\lq hard-scattering approach \rq\rq,
emerges if one
tries to figure out the leading twist contributions to hard exclusive processes
within
perturbative QCD~\cite{BL89}. The outcome of such an analysis is a
factorization formula which
is also expressed as a convolution integral. This integral now consists of
distribution
amplitudes (DAs) and a hard scattering amplitude. The process dependent
hard scattering
amplitude is perturbatively calculable and represents the scattering of the
hadronic
constituents in collinear approximation. The process independent DAs contain
the
non-perturbative bound-state dynamics of the hadronic constituents. DAs are,
roughly speaking,
valence Fock-state wave functions integrated over the transverse momentum. At
present, the
knowledge on hadron DAs is still rather limited. The main information is
provided by QCD
sum-rule techniques which give estimates of the lowest moments  of various
meson and baryon
DAs \cite{CZ84,FZOZ88,COZ89}. The few lowest moments impose some restrictions
on the shape of the DAs
but do not determine them uniquely. A thorough discussion on how to construct
model DAs
reproducing a certain number of moments can be found in Ref.~\cite{St94}.
One should also note that the DAs constrained by QCD sum rules are
subject to severe criticism~\cite{JKR96,Bo95,Br94,Ra91}.
Thus one is forced,
at present, to supplement the lack of theoretical knowledge on DAs by some
input coming from
experiment. Photoproduction reactions are, in this respect, certainly very
interesting. They
exhibit a rich flavour structure and are still simple enough to allow for the
computation of
all the  Feynman diagrams which enter the hard scattering amplitude.
Perturbative QCD
predictions for various photoproduction channels have been published in
Ref.~\cite{FHZ91}. This
paper discusses also the sensitivity of the results on the choice of the hadron
DAs. Apart from
the fact that there are objections to the numerics of this work (cf.
Sect.~\ref{sec:results}) and
it still needs confirmation, the predictions for the K-$\Lambda$ channel occur
to be in
considerable disagreement with experiment.

The present work concentrates on photoproduction of the K-$\Lambda$ and
K$^{\ast}$-$\Lambda$
final states.~We consider these reactions within a particular
version of the hard-scattering approach in
which baryons are treated as quark-diquark systems. The same approach has
already been applied
successfully to other photon-induced hadronic reactions like magnetic
and electric baryon form factors in
the space- \cite{Ja93} and time-like region \cite{Kro93}, real and virtual
Compton scattering
\cite{Kro95}, and two-photon annihilation into proton-antiproton \cite{Kro93}.
Further applications
of the diquark model include the charmonium decay $\eta_{\text{c}}
\rightarrow \text{p}
\bar{\text{p}}$ \cite{Kro93} and the calculation of Landshoff contributions in
elastic proton-proton
scattering \cite{Ja94}. The introduction of diquarks does not only simplify
computations, but is
rather motivated by the requirement to extend the hard-scattering approach from
large down to
intermediate momentum transfers ($p_{\perp}^2 \ageq 4$~GeV$^2$). This is the
momentum-transfer
region where experimental data are still available, but where still persisting
non-perturbative
effects prevent the pure quark hard-scattering approach to become fully
operational. Diquarks may be
considered as an effective way to cope with such non-perturbative effects. It
is an assumption that, on
an intermediate momentum-transfer scale, two of the three valence quarks in a
baryon make up a diquark cluster. However,~from many experimental
and theoretical approaches there have been indications suggesting
the presence of diquarks. For instance, they were introduced in baryon
spectroscopy, in nuclear physics, in jet fragmentation and in weak
interactions to explain the famous $\Delta I=1/2$~rule. Diquarks also
provide a natural explanation of the equal slopes of meson and baryon
Regge trajectories. For more details and for references, see
\cite{Kro:87}. It is important to note that QCD provides some
attraction between two quarks in a colour $\{\bar{3}\}$ state at
short distances as is to be seen from the static reduction of the
one-gluon exchange term. Also the instanton
force seem to lead to diquark formation \cite{shu:84}. Even more
important for our aim, diquarks have also been found to play a role in
inclusive hard-scattering reactions. The most obvious place to signal
their presence is deep inelastic lepton-nucleon scattering. Indeed
the higher twist terms, convincingly observed \cite{Vir:91}, can be
modeled as lepton-diquark elastic scattering. Baryon production in
inclusive p-p collisions also clearly reveals the need for diquarks
scattered elastically in the hard elementary reactions
\cite{szc:92}. For instance, kinematical dependencies or the excess of
the proton yield over the antiproton yield find simple explanations
in the diquark model. No other explanation of these phenomena is
known as yet.

The main ingredients of the diquark model are baryon DAs in terms of quarks and
diquarks, the
coupling of gluons and photons to diquarks, and, in order to account for the
composite nature of
diquarks, phenomenological diquark form factors. The proper choice of the
diquark form factors
guarantees the compatibility of the diquark model with the pure quark
hard-scattering approach in the
limit $p_{\perp} \rightarrow \infty$. In so far
the pure quark picture of Brodsky-Lepage and the diquark model do not
oppose each other, they are not alternatives but rather complements.
The model parameters have been determined in
Ref.~\cite{Ja93} by means of elastic electron-nucleon scattering data.  The
full model
incorporates scalar (S) and vector diquarks. Vector diquarks are important for
the description
of spin observables which violate hadronic helicity conservation, i.e.
quantities not explicable
within the pure quark hard-scattering approach. The nice and simplifying
feature
of the two
photoproduction reactions we are interested in is that they are not influenced
by vector diquarks
since only the S$_{\left[ \text{u,d} \right]}$ diquark is common to proton and
$\Lambda$.

The following section starts with an outline of the hard-scattering approach
with (scalar)
diquarks. It contains also a description of the Kaon, proton, and $\Lambda$ DAs
to be used in the
sequel. Section~\ref{sec:photoprod} deals with the constituent kinematics,
photoproduction
observables, and the general structure of photoproduction helicity amplitudes
within the diquark
model. Predictions for photoproduction observables with a discussion of their
dependence on the
choice of the Kaon DA can be found in Sec.~\ref{sec:results}. Conclusions and
prospects are given in
Sec.~\ref{sec:summary}. Analytical expressions for the helicity amplitudes are
tabulated in
Appendix~\ref{sec:eamp}. Our numerical method for treating propagator
singularities is sketched in
Appendix~\ref{sec:proppol}.

\section{Hard Scattering with Diquarks}
\label{sec:diquark}
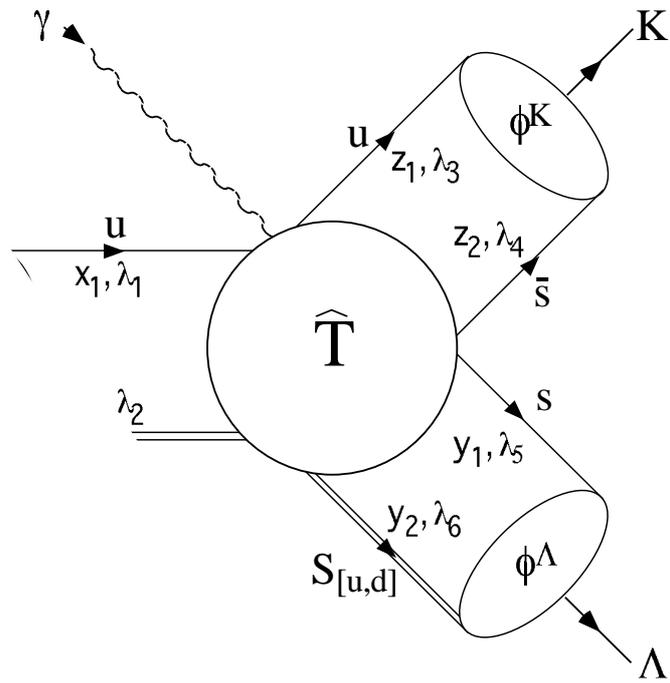
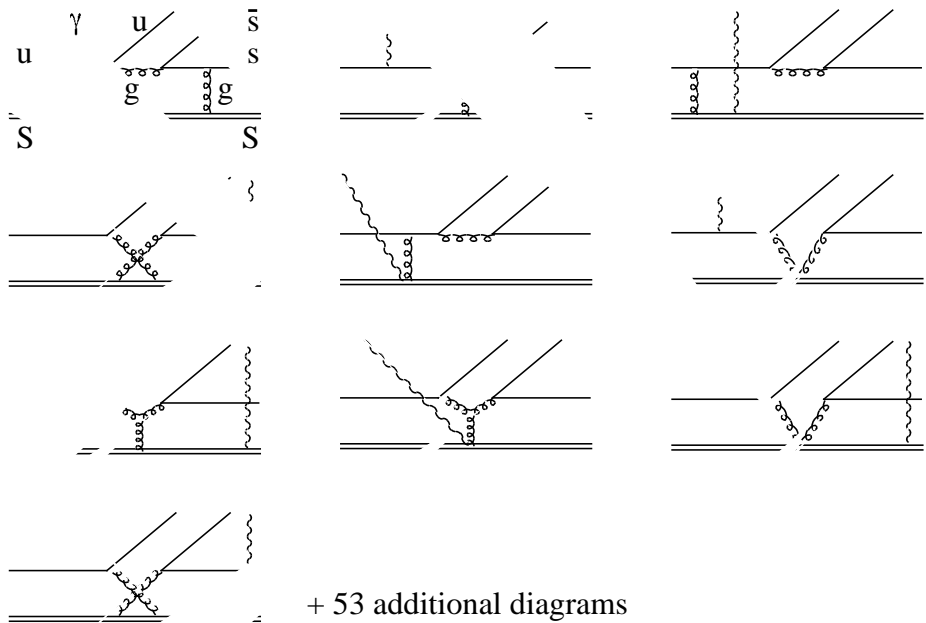
\begin{figure}
\caption{A few representative examples of Feynman diagrams contributing to the
elementary process $\gamma \text{u} \text{S}_{\left[ \text{u,d} \right]}
\longrightarrow \text{u}
\bar{\text{s}}
\text{s} \text{S}_{\left[ \text{u,d} \right]}$.}
\label{feyn}
\end{figure}

Within the hard-scattering approach a helicity amplitude $M_{\{\lambda\}}$ for
the reaction
$\gamma \text{p} \longrightarrow \text{K}^{(\ast)+} \Lambda$ is (to leading
order in
$1/p_\perp$) given by the the convolution integral \cite{BL89}
\widetext
\begin{equation}
M_{\{\lambda\}}(\hat{s},\hat{t}) = \int_0^1 dx_1 dy_1 dz_1
{\phi^{\text{K}^{(\ast)}}}^{\dagger}(z_1,\tilde{Q})
{\phi^{\Lambda}}^{\dagger}(y_1,\tilde{Q})
\widehat{T}_{\{\lambda\}}(x_1,y_1,z_1;\hat{s},\hat{t})
\phi^{\text{p}}(x_1,\tilde{Q}) \, . \label{convol}
\end{equation}
\narrowtext
\noindent
The distribution amplitudes $\phi^{\text{H}}$ are probability amplitudes for
finding the
valence Fock state in the hadron H with the constituents carrying certain
fractions of the
momentum of their parent hadron and being  collinear up to a factorization
scale $\tilde{Q}$.
In our model the valence Fock state of an ordinary baryon is assumed to consist
of a
quark and a diquark (D). We fix our notation in such a way that the momentum
fraction
appearing in the argument of $\phi_{\text{H}}$ is carried by the quark -- with
the momentum
fraction of the other constituent (either diquark or antiquark) it sums up to 1
(cf.
Fig.~\ref{kinem}). In what follows we will neglect the (logarithmic)
$\tilde{Q}$ dependence of
the DAs since it is of minor importance in the restricted energy range we will
be interested
in. The hard scattering amplitude $\widehat{T}_{\{\lambda\}}$ is calculated
perturbatively in
collinear approximation and consists in our particular case of all possible
tree diagrams
contributing to the elementary scattering process $\gamma \text{u} \text{S}
\longrightarrow
\text{u} \bar{\text{s}} \text{s} \text{D}$. A few examples of such diagrams are
depicted in
Fig.~\ref{feyn}. The subscript ${\{\lambda\}}$ represents the set of possible
photon, proton
and $\Lambda$ helicities. We have written the Mandelstam variables $s$ and $t$
with a hat to
indicate that masses are neglected during the calculation of the hard
scattering amplitude.
They are only taken into account in flux and phase-space factors.

If one assumes zero relative orbital angular momentum between quark and diquark
and takes
advantage of the collinear approximation ($p_{\text{q}} = x_1 p_{\text{B}}$ and
$p_{\text{D}} = x_2 p_{\text{B}} = (1 - x_1) p_{\text{B}}$) the valence
Fock-state
wave function of a baryon B belonging to the energetically lowest lying octet
may be written as
\widetext
\begin{equation}
\Psi^{\text{B}} ( p_{\text{B}}; \lambda) =
f_{\text{S}}^{\text{B}}
\phi_{\text{S}}^{\text{B}} (x_1) \, \chi^{\text{B}}_{\text{S}} \,
u( p_{\text{B}},\lambda)
 + f_{\text{V}}^{\text{B}} \, \phi_{\text{V}}^{\text{B}} (x_1) \,
 \chi^{\text{B}}_{\text{V}} \, \frac{ 1 }{ \sqrt{3}} \, (\gamma^\alpha +
\frac{p_{\text{B}}^\alpha }{ m_{\text{B}} }) \, \gamma_5 u(
p_{\text{B}},\lambda )
\, .
\label{wfcov}
\end{equation}
The two terms in Eq.\ (\ref{wfcov}) represent configurations consisting of a
quark and
either a scalar or vector diquark. The pleasant feature of the covariant
wave-function
representation Eq.(\ref{wfcov}) is that it contains, besides $x_1$ and $\alpha$
(the Lorentz index
of the vector-diquark polarization vector), only baryonic quantities (momentum
$p_{\text{B}}$,
helicity $\lambda$, baryon mass $m_{\text{B}}$).

For an SU(6)-like spin-flavour dependence the flavour functions $\chi$ for
proton and $\Lambda$ take on the form (the notation should be obvious)
\begin{equation}
\chi_{\text{S}}^{\text{p}} = \text{u} \text{S}_{\text{[u,d]}} \, , \;\;\;
\chi_{\text{V}}^{\text{p}} = \phantom{-} [\text{u} \text{V}_{\text{\{u,d\}}}
-\sqrt{2} \text{d}
\text{V}_{\text{\{u,u\}}}] / \sqrt{3} \, ,
\label{flavp}
\end{equation}
\begin{equation}
\chi_{\text{S}}^{\Lambda} = [ \text{u} \text{S}_{\text{[d,s]}} - \text{d}
\text{S}_{\text{[u,s]}} - 2 \text{s} \text{S}_{\text{[u,d]}}] /
\sqrt{6}
\, ,
\;\;\;
\chi_{\text{V}}^{\Lambda} =  [\text{u} \text{V}_{\text{\{d,s\}}}- \text{d}
\text{V}_{\text{\{u,s\}}}] / \sqrt{2} \, .
\label{flavl}
\end{equation}

Similarly, also the q-$\bar{\text{q}}$ wave functions of pseudoscalar (PM) and
vector (VM)
mesons may be represented in a covariant way
\begin{equation}
\Psi^{\text{PM}} ( p_{\text{PM}}) =
f^{\text{PM}}  \, \phi^{\text{PM}} (x_1) \, \chi^{\text{PM}} \,
\frac{1}{ \sqrt{2}}(\not{p}_{\text{PM}} +  m_{\text{PM}} ) \gamma_5
\, , \label{wfcovps}
\end{equation}
\begin{equation}
\Psi^{\text{VM}} ( p_{\text{VM}}; \lambda) =
- f^{\text{VM}}  \, \phi^{\text{VM}} (x_1,\lambda) \, \chi^{\text{VM}} \,
\frac{1}{ \sqrt{2}}(\not{p}_{\text{VM}} +  m_{\text{VM}} )
\not{\epsilon}(\lambda)
\, , \label{wfcovv}
\end{equation}
\narrowtext
\noindent
with the flavour function of the K$^{(\ast) +}$ meson given by
\begin{equation}
\chi^{{\text{K}^{(\ast)}}^{+}} = \text{u} \bar{\text{s}} \, .
\end{equation}

At this point we are already in the position to recognize a considerable
simplification
in the treatment of the reaction $\gamma \text{p} \longrightarrow
{\text{K}^{(\ast)}}^{+} \Lambda$
as compared to arbitrary photoproduction processes. Photoproduction of the
${\text{K}^{(\ast)}}^{+}$-$\Lambda$ final state can solely proceed via the
$\text{S}_{\text{[u,d]}}$
diquark. This is the only kind of diquark occurring in both, the proton and the
$\Lambda$
wave function (cf. Eqs.\ (\ref{flavp}) and (\ref{flavl})). The opposite
situation, namely that
only the $\text{V}_{\text{ \{ u,d \} }}$ diquark becomes involved, holds for
$\gamma\text{p}
\longrightarrow {\text{K}^{(\ast)}}^{+} \Sigma^0$. The fact that scalar
diquarks as well as (massless)
quarks do not change their helicity when interacting with a gluon imposes
already strong
restrictions on spin observables of the $\text{K}^{+}$-$\Lambda$ and
${\text{K}^{\ast}}^{+}$-
$\Lambda$ channels. Helicity amplitudes which require the flip of the baryonic
helicity are
predicted to vanish, e.g., for the $\gamma \text{p} \longrightarrow
\text{K}^{+} \Lambda$ process.
On the other hand, helicity flips may take place in the  $\gamma\text{p}
\longrightarrow \text{K}^{+}
\Sigma^0$ reaction by means of the vector diquark. In order to work out the
different features of
scalar and vector diquarks a comparison of the $\Lambda$ and $\Sigma^0$
photoproduction
channels would certainly be of great benefit.

The complicated, non-perturbative bound-state dynamics is contained in the DAs
$\phi^{\text{H}}$.
These are light-cone wave functions integrated over transverse momentum (up to
$\tilde{Q}$). The $r=0$ values of the corresponding configuration space wave
functions are related to
the constants $f^{\text{H}}$. We will check the sensitivity of our
photoproduction calculation
on the shape of the K (K$^{\ast}$) DA by choosing two qualitatively rather
different forms: the
one is the asymptotic DA
\begin{equation}
\phi_{\text{asy}} (x) = 6 x (1-x)  \; ,
\label{DAKa}
\end{equation}
which solves the $\tilde{Q}$ evolution equation for $\phi(x,\tilde{Q})$ in the
limit $\tilde{Q}
\rightarrow \infty$ (see, e.g., Ref.~\cite{BL89}); the other one is a
two-humped DA, namely
\widetext
\begin{equation}
\phi^{\text{K}}_{\text{CZ}} (x) = N^{\text{K}} \phi_{\text{asy}} (x)
[ 0.08 + 0.6 (1 - 2 x)^2 + 0.25 (1 - 2 x)^3 ]
\; ,
\label{DAKCZ}
\end{equation}
for the K,
\begin{equation}
\phi^{\text{K}^{\ast}}_{\text{L}} (x) = N^{\text{K}^{\ast}}_{\text{L}}
\phi_{\text{asy}} (x)
[ 0.18 + 0.1 (1 - 2 x)^2 + 0.41 (1 - 2 x)^3 ]  \; ,
\label{DAKsCZl}
\end{equation}
for the longitudinally polarized K$^{\ast}$, and
\begin{equation}
\phi^{\text{K}^{\ast}}_{\text{T}} (x)  =  N^{\text{K}^{\ast}}_{\text{T}}
\phi_{\text{asy}} (x)
[ 0.284 +0.07 (1 - 2 x) - 0.534 (1 - 2 x)^2 +
0.21 (1 - 2 x)^3 + 0.267 (1 - 2 x)^4]  \; ,
\label{DAKsCZt}
\end{equation}
\narrowtext
\noindent
for the transversially polarized K$^{\ast}$. The DAs,
Eqs.~(\ref{DAKCZ})-(\ref{DAKsCZt}), have been
proposed in Refs.~\cite{CZ84} and \cite{BC90}. They reproduce the corresponding
QCD sum-rule moments
at $\tilde{Q}^2 = 0.25$~GeV$^2$. It has been demonstrated quite recently that
the linear
$x$-dependence of pseudoscalar meson DAs at the end points $x \rightarrow 0,1$
can be considered as
a direct consequence of QCD \cite{ChZ95}.

The usual normalization condition, $\int_0^1 dx \phi^{\text{H}}(x) = 1$, fixes
the constants $N$
in Eqs.~(\ref{DAKCZ})-(\ref{DAKsCZt}). The quantities $f^{\text{PM}}$ and
$f^{\text{VM}}$ showing up
in Eqs.\ (\ref{wfcovps}) and (\ref{wfcovv}) are related to experimentally
determinable decay constants
of the corresponding mesons. From the $\text{K}^+ \rightarrow \mu^+ \nu_{\mu}$
decay one infers in
particular that $f^{\text{K}^+} = f^{\text{K}^+}_{\text{decay}}/{2 \sqrt{6}} =
32.6$ MeV. The value
of $f^{\text{K}^{\ast}}$ is only known indirectly via the QCD-sum-rule result
$f^{\text{K}^{\ast}} = 1.05 f^{\rho}$ (cf. Ref.~\cite{CZ84}). The experimental
value of $f^{\rho}
= f^{\rho}_{\text{decay}}/{2 \sqrt{6}} = 40.8$ MeV obtained from the $\rho^0
\rightarrow \text{e}^+
\text{e}^-$ decay implies $f^{\text{K}^{\ast}} = 42.9$ MeV.

In previous applications
of the diquark model \cite{Ja93,Kro93,Kro95} a DA of the form
\begin{equation}
\phi_{\text{S}}^{\text{B}} (x) = N_{\text{S}} x (1-x)^3
\exp \left[ - b^2 \left( \frac{m_{\text{q}}^2}{x} + \frac{m_{\text{S}}^2 }{
(1-x)} \right) \right] \;
, \label{DAp}
\end{equation}
proved to be quite appropriate for the quark-scalar diquark Fock state of octet
baryons B. The origin of the DA, Eq.~(\ref{DAp}), is a nonrelativistic
harmonic-oscillator wave function~\cite{Hua89}. Therefore the masses
appearing in the exponentials have to be considered as
constituent masses (330 MeV for light quarks, 580 MeV for light diquarks,
strange quarks are 150 MeV heavier than light quarks). The oscillator
parameter $b^2 = 0.248$ GeV$^{-2}$ is chosen in such
a way that the full wave function gives rise to a value of $600$ MeV for the
mean intrinsic transverse momentum of quarks inside a nucleon. Note, that the
DA,
Eq.~(\ref{DAp}), exhibits a flavour dependence due to the masses in the
exponential.
The exponential in (\ref{DAp}) is merely introduced for theoretical
purposes (e.~g.~in order to suppress the soft end-point regions). In
the actual data fitting the exponential plays only a minor
role. Therefore, the masses and the oscillator parameter are not
considered as free parameters but taken from the literature. We stress
that the constituent masses do not appear in the hard scattering amplitudes.

The dynamics of diquarks is governed by their coupling to gluons and photons.
With respect to colour the diquark behaves like an antiquark. In order that the
diquark
in combination with a colour-triplet quark gives a colourless baryon it has to
be in
a colour antitriplet state. The colour part of the quark-diquark wave function
(omitted in
Eq.~(\ref{wfcov}) is therefore $\psi_{\text{qD}}^{\text{colour}} = (1/\sqrt{3})
\sum_{a=1}^3
\delta_{a \bar{a}}$. The Feynman rules of electromagnetically interacting
scalar diquarks are just
those of standard scalar electrodynamics \cite{BD65}. Replacement of the
electric charge
$e_0 e_{\text{S}}$ by $-g_{\text{s}} t^a$, with $g_{\text{s}}=\sqrt{4 \pi
\alpha_{\text{s}}}$ denoting
the  strong coupling constant and $t^a = \lambda^a/2$ Gell-Mann colour
matrices, yields the
corresponding Feynman rules for strongly interacting scalar diquarks. The
explicit expressions for
$\gamma$-S and g-S vertices read:
\begin{eqnarray}
\text{S} \gamma \text{S}: &  - i e_0 e_{\text{S}} (p_1 + p_2)_{\mu} \; , \qquad
&\gamma \text{S} \text{g} \text{S}:  - 2 i e_0 e_{\text{S}} g_{\text{s}} t^a
g_{\mu \nu} \; ,
\nonumber \\
\text{S} \text{g} \text{S}: & \phantom{-} i g_{\text{s}} t^a (p_1 + p_2)_{\mu}
\; , \qquad
&\text{g} \text{S} \text{g} \text{S}:  i g_{\text{s}}^2 \{t^a,t^b\} g_{\mu \nu}
\; . \nonumber \\
\end{eqnarray}
During the calculation of Feynman diagrams diquarks are treated as point-like
particles.
The composite nature of diquarks is taken into account by multiplying the
expressions for the various
Feynman diagrams with diquark form factors
\begin{equation}
F_{\text{S}}^{(n+2)}(Q^2) = \delta_{\text{S}}
\frac{Q_{\text{S}}^2}{Q_{\text{S}}^2 + Q^2}
\left\{
\begin{array}{ll} 1 & n = 1 \\ a_{\text{S}} & n \geq 2 \end{array} \right.
\end{equation}
which depend on the number ($n$) of gauge bosons going to the diquark. This
choice of the form
factors ensures that the scaling behaviour of the diquark model goes over into
that of the pure quark
model in the limit $p_{\perp} \rightarrow \infty$. The factor
$\delta_{\text{S}}  =
\alpha_{\text{s}} (Q^2) / \alpha_{\text{s}} (Q^2_{\text{S}})$
($\delta_{\text{S}} = 1$ for $Q^2 \leq
Q^2_{\text{S}}$) provides the correct powers of $\alpha_{\text{s}} (Q^2)$ for
asymptotically large
$Q^2$. For the running coupling constant $\alpha_{\text{s}}$ the one-loop
result
$\alpha_{\text{s}} = 12 \pi / 25 \ln (Q^2 / \Lambda_{\text{QCD}}^2 )$
is used with $\Lambda_{\text{QCD}} = 200$ MeV. In addition, $\alpha_{\text{s}}$
is restricted
to be smaller than $0.5$. The possibility of diquark excitation and break-up in
intermediate states
where diquarks can be far off-shell is taken into consideration by means of the
strength parameter
$a_{\text{S}}$.

Due to the reasons already mentioned, vector diquarks do not show up in the
reactions we are
investigating in the present paper. However, they have been dealt with in
Refs.~\cite{Ja93,Kro93,Kro95} to which we refer for further details of the
diquark model.

\section{Photoproduction of Mesons\ -- \protect \\ Kinematics and Helicity
Amplitudes}
\label{sec:photoprod}
Exclusive photoproduction of pseudoscalar mesons can, in general, be described
by 4 independent
helicity amplitudes. Following the notation of Ref.~\cite{Ba75} we denote these
amplitudes by
\begin{eqnarray}
N = & M_{0,-\frac{1}{2},+1,+\frac{1}{2}} \, , \qquad S_1 = &
M_{0,-\frac{1}{2},+1,-\frac{1}{2}} \, ,
\nonumber \\
D = & M_{0,+\frac{1}{2},+1,-\frac{1}{2}} \, , \qquad S_2 = &
M_{0,+\frac{1}{2},+1,+\frac{1}{2}} \, .
\end{eqnarray}
$N$ , $S_1$, $S_2$, and $D$ represent non-flip, single-flip, and double-flip
amplitudes,
respectively. Our helicity amplitudes are normalized in such a way that the
unpolarized differential
cross section is given by
\begin{equation}
\frac{d\sigma}{dt} = \frac{1}{32 \pi (s - m^2_{\text{p}})^2}
\left( \vert N \vert^2 + \vert S_1 \vert^2 + \vert S_2 \vert^2 + \vert D
\vert^2 \right) \, .
\label{unpolwq}
\end{equation}
As we have argued above two of the four amplitudes vanish, $N = D = 0$, if we
concentrate on the
particular process $\gamma \text{p} \longrightarrow \text{K}^{+} \Lambda$ and
treat it within the
diquark model. Out of the 15 polarization observables discussed in
Ref.~\cite{Ba75} there are only
3 observables which remain nonzero and which differ from each other (and from
$d\sigma/dt$) for
vanishing $N$ and $D$. These can be chosen as the photon asymmetry
\begin{equation}
\Sigma \frac{d\sigma}{dt} = \frac {d\sigma_{\perp}}{dt} -
\frac{d\sigma_{\|}}{dt} =
\frac{1}{16 \pi (s - m^2_{\text{p}})^2}
\Re \left( S_1^{\ast} S_2 - N D^{\ast} \right) \, ,
\end{equation}
and the two double-polarization observables
\begin{equation}
G \frac{d\sigma}{dt} = - \frac{1}{16 \pi (s - m^2_{\text{p}})^2}
\Im \left( S_1 S_2^{\ast} + N D^{\ast} \right) \, ,
\end{equation}
and
\begin{equation}
E \frac{d\sigma}{dt} = \frac{1}{32 \pi (s - m^2_{\text{p}})^2}
\left( \vert N \vert^2 - \vert S_1 \vert^2 + \vert S_2 \vert^2 - \vert D
\vert^2 \right) \, .
\end{equation}
$d\sigma_{\perp}$ ($d\sigma_{\|}$) denotes the cross section for photons
polarized perpendicular
(parallel) to the reaction plane.

Photoproduction of vector mesons may be expressed by altogether 12 linear
independent helicity
amplitudes \cite{Ta94}. The diquark model leaves four amplitudes nonzero if
applied to the formation
of the ${\text{K}^{\ast}}^{+}$-$\Lambda$ final state. In addition to $S_1$ and
$S_2$ one
can choose, e.g., $M_{+1,-\frac{1}{2},+1,+\frac{1}{2}}$ and
$M_{-1,+\frac{1}{2},+1,-\frac{1}{2}}$.
Due to the lack of experimental data we will restrict our discussion of
${\text{K}^{\ast}}^{+}$-$\Lambda$ production to the unpolarized differential
cross section which is
obtained from Eq.~(\ref{unpolwq}) by including also $\vert
M_{+1,-\frac{1}{2},+1,+\frac{1}{2}}
\vert^2$ and
$\vert M_{-1,+\frac{1}{2},+1,-\frac{1}{2}} \vert^2$.

The hard scattering amplitude $\widehat{T}_{\{\lambda\}}$ for the elementary
process $\gamma \text{u}
\text{S}_{\text{[u,s]}} \longrightarrow \text{u} \bar{\text{s}} \text{s}
\text{S}_{\text{[u,s]}}$
consists, in general, of 79 different tree diagrams. However, only 63 diagrams
are encountered
if the outgoing meson is a $\text{K}^+$ or ${\text{K}^{\ast}}^+$. The other 16
diagrams
require the $\text{s}$-$\bar{\text{s}}$ pair to go into the produced meson.
Diagrams of this type
would, e.g., be important in $\gamma \text{p} \longrightarrow \phi \text{p}$.

The helicity structure of the hard scattering amplitude
$\widehat{T}_{\{\lambda\}}$ is particularly
simple for the $\text{K}^+$-$\Lambda$ and ${\text{K}^{\ast}}^+$-$\Lambda$ final
states. Assigning
helicity labels to the hadronic constituents as in Fig.~\ref{kinem} one finds
(with the S diquark
helicities $\lambda_2 = \lambda_6 = 0$):
\begin{eqnarray}
\lambda_{\text{p}} & = \lambda_1 & =  \phantom{-} \lambda_3  \, , \nonumber \\
\lambda_{\Lambda} & =  \lambda_5 & = - \lambda_4 \, .
\label{helconda}
\end{eqnarray}
Thus the quark helicities are uniquely determined by the proton and $\Lambda$
helicity, respectively.
The additional relation (hadronic helicity conservation)
\begin{equation}
\lambda_3  + \lambda_4  =  \lambda_{\text{p}} - \lambda_{\Lambda} =
\lambda_{\text{K}^{(\ast)}}
\label{helcondb}
\end{equation}
is the condition for the hard scattering amplitude $\widehat{T}_{\{\lambda\}}$
and consequently the
hadronic amplitude $M_{\{\lambda\}}$ to become nonzero within the diquark
model.

 As depicted in Fig.~\ref{feyn} the hard scattering amplitude
$\widehat{T}_{\{\lambda\}}(x_1,y_1,z_1;\hat{s},\hat{t})$ for the elementary
process $\gamma
\text{u} \text{S}_{\left[ \text{u,d} \right]} \longrightarrow \text{u}
\bar{\text{s}} \text{s}
\text{S}_{\left[ \text{u,d} \right]}$  can be decomposed into 3-, 4-, and
5-point contributions
\widetext
\begin{eqnarray}
\widehat{T}_{\{\lambda\}}(x_1,y_1,z_1;\hat{s},\hat{t}) & = & -
\frac{2}{\sqrt{6}}
e_{\text{u\phantom{d}}} \left(
\widehat{T}_{\{\lambda\}}^{(3,\text{q})}(x_1,y_1,z_1;\hat{s},\hat{t}) +
\widehat{T}_{\{\lambda\}}^{(4,\text{q})}(x_1,y_1,z_1;\hat{s},\hat{t}) \right)
\nonumber \\
&  & - \frac{2}{\sqrt{6}} e_{\text{s\phantom{d}}} \left(
\widehat{T}_{\{\lambda\}}^{(3,\bar{\text{q}})}(x_1,y_1,z_1;\hat{s},\hat{t}) +
\widehat{T}_{\{\lambda\}}^{(4,\bar{\text{q}})}(x_1,y_1,z_1;\hat{s},\hat{t})
\right) \nonumber \\
&  & - \frac{2}{\sqrt{6}} e_{\text{ud}} \left(
\widehat{T}_{\{\lambda\}}^{(4,\text{S})}(x_1,y_1,z_1;\hat{s},\hat{t}) +
\widehat{T}_{\{\lambda\}}^{(5,\text{S})}(x_1,y_1,z_1;\hat{s},\hat{t}) \right)
\, ,
\label{Tn}
\end{eqnarray}
depending on whether one, two, or three gauge bosons go to the diquark. The
additional
superscripts q, $\bar{\text{q}}$, and S occurring in Eq.~(\ref{Tn}) indicate
whether the
photon couples to the u quark, the s quark, or the S diquark, respectively. For
the
numerical evaluation of the convolution integral Eq.~(\ref{convol}) it is
advantageous to
further subdivide the various n-point contributions into two parts which differ
by their
propagator singularities:
\begin{eqnarray}
\widehat{T}_{\{\lambda\}}^{(3,\text{q})}(x_1,y_1,z_1;\hat{s},\hat{t})
& = & \frac{f_{\{\lambda\}}^{(3,\text{q})}(x_1,y_1,z_1;\hat{s},\hat{t})}
           {(q_2^2 + i \epsilon) (g_1^2 + i \epsilon^\prime)}
  +   \frac{g_{\{\lambda\}}^{(3,\text{q})}(x_1,y_1,z_1;\hat{s},\hat{t})}
           {(q_3^2 + i \epsilon) (q_4^2 + i \epsilon^\prime)} \, , \nonumber \\
\widehat{T}_{\{\lambda\}}^{(3,\bar{\text{q}})}(x_1,y_1,z_1;\hat{s},\hat{t})
& = & \frac{f_{\{\lambda\}}^{(3,\bar{\text{q}})}(x_1,y_1,z_1;\hat{s},\hat{t})}
           {(q_2^2 + i \epsilon) (q_5^2 + i \epsilon^\prime)}
  +   \frac{g_{\{\lambda\}}^{(3,\bar{\text{q}})}(x_1,y_1,z_1;\hat{s},\hat{t})}
           {(q_3^2 + i \epsilon) (g_3^2 + i \epsilon^\prime)} \, , \nonumber \\
\widehat{T}_{\{\lambda\}}^{(4,\text{q})}(x_1,y_1,z_1;\hat{s},\hat{t})
& = & \frac{f_{\{\lambda\}}^{(4,\text{q})}(x_1,y_1,z_1;\hat{s},\hat{t})}
           {(g_1^2 + i \epsilon) (D_1^2 + i \epsilon^\prime)}
  +   \frac{g_{\{\lambda\}}^{(4,\text{q})}(x_1,y_1,z_1;\hat{s},\hat{t})}
           {(g_1^2 + i \epsilon)} \, , \nonumber \\
\widehat{T}_{\{\lambda\}}^{(4,\bar{\text{q}})}(x_1,y_1,z_1;\hat{s},\hat{t})
& = & \frac{f_{\{\lambda\}}^{(4,\bar{\text{q}})}(x_1,y_1,z_1;\hat{s},\hat{t})}
           {(g_3^2 + i \epsilon) (D_2^2 + i \epsilon^\prime)}
  +   \frac{g_{\{\lambda\}}^{(4,\bar{\text{q}})}(x_1,y_1,z_1;\hat{s},\hat{t})}
           {(g_3^2 + i \epsilon)} \, , \nonumber \\
\widehat{T}_{\{\lambda\}}^{(4,\text{S})}(x_1,y_1,z_1;\hat{s},\hat{t})
& = & \frac{f_{\{\lambda\}}^{(4,\text{S})}(x_1,y_1,z_1;\hat{s},\hat{t})}
           {(q_5^2 + i \epsilon) (g_2^2 + i \epsilon^\prime)}
  +   \frac{g_{\{\lambda\}}^{(4,\text{S})}(x_1,y_1,z_1;\hat{s},\hat{t})}
           {(q_4^2 + i \epsilon) (g_2^2 + i \epsilon^\prime)} \, , \nonumber \\
\widehat{T}_{\{\lambda\}}^{(5,\text{S})}(x_1,y_1,z_1;\hat{s},\hat{t})
& = & \frac{f_{\{\lambda\}}^{(5,\text{S})}(x_1,y_1,z_1;\hat{s},\hat{t})}
           {(D_1^2 + i \epsilon) (D_2^2 + i \epsilon^\prime)}
  +   \frac{g_{\{\lambda\}}^{(5,\text{S})}(x_1,y_1,z_1;\hat{s},\hat{t})}
           {(g_2^2 + i \epsilon)} \, .
\label{Tfg}
\end{eqnarray}
Apart from $g_3^{-2}$ the $q_i^{-2}$, $D_i^{-2}$, and $g_i^{-2}$ denote just
those quark,
diquark, and gluon propagators which can go on-shell when integrating over
$x_1$, $y_1$, and
$z_1$. In order to make symmetry properties of the functions $f$ and $g$ with
respect to
interchange of Mandelstam variables and momentum fractions more obvious (cf.
Appendix\
\ref{sec:eamp}) we have also extracted the non-singular gluon propagator
$g_3^{-2}$. Explicitly
the propagator denominators read:
\begin{eqnarray}
q_2^2 & = y_2 z_2 \hat{s} + x_2 y_2 \hat{t} + x_2 z_2 \hat{u} \, , \qquad
g_1^2 & = z_2 \hat{s} + x_2 \hat{t} + x_2 z_2 \hat{u} \, , \nonumber \\
q_3^2 & = y_2 z_1 \hat{s} + x_2 y_2 \hat{t} + x_2 z_1 \hat{u} \, , \qquad
g_2^2 & = y_1 \hat{s} + x_1 y_1 \hat{t} + x_1 \hat{u} \, , \nonumber \\
q_4^2 & = y_1 z_2 \hat{s} + x_1 y_1 \hat{t} + x_1 z_2 \hat{u} \, , \qquad
g_3^2 & = y_2 z_1 \hat{s} + y_2 \hat{t} + z_1 \hat{u} \, , \nonumber \\
q_5^2 & = y_1 z_1 \hat{s} + x_1 y_1 \hat{t} + x_1 z_1 \hat{u} \, , \qquad
D_1^2 & = y_1 z_2 \hat{s} + x_2 y_1 \hat{t} + x_2 z_2 \hat{u} \, , \nonumber \\
      &\phantom{ = y_1 z_1 \hat{s} + x_1 y_1 \hat{t} + x_1 z_1 \hat{u} \, ,}
\qquad
D_2^2 & = y_2 z_1 \hat{s} + x_1 y_2 \hat{t} + x_1 z_1 \hat{u} \, .
\label{prop}
\end{eqnarray}
\narrowtext
\noindent
As already indicated in Eq.~(\ref{Tfg}) propagator singularities are treated by
means of
the usual $i \epsilon$ prescription.

Analytical expressions for the functions $f$ and $g$ (cf.
Appendix~\ref{sec:eamp}) have been
derived with the help of \lq\lq FeynArts\rq\rq  \cite{Kue90} and \lq\lq
FeynCalc\rq\rq \cite{Me91}
-- two program packages written in \lq\lq Mathematica \rq\rq which serve the
automatic generation and
evaluation of Feynman diagrams. Since \lq\lq FeynArts\rq\rq\ only contains the
Feynman rules of the
Standard Model it had to be extended to deal with S diquarks as well. The
spinor techniques developed
by Kleiss and Stirling \cite{Kle85} have been utilized to convert strings of
gamma matrices sandwiched
between spinors into traces which can be handled by \lq\lq FeynCalc\rq\rq. A
strong indication for
the correctness of our results is already the agreement (apart from a detected
sign error) with a
previous independent calculation \cite{Schue92} performed with \lq\lq
FORM\rq\rq  \cite{Ver91}
(another symbolic computer program for high-energy physics). Further checks of
our analytical results
were carried out by testing the U(1) gauge invariance with respect to the
photon and the SU(3) gauge
invariance with respect to the gluon. The proof of gauge invariance is
facilitated by observing
that not only the sum of all 63 tree diagrams gives a gauge invariant
expression, but rather each of
the functions $f$ and $g$ in Eq.~(\ref{Tfg}) is by itself gauge invariant. In
addition to the gauge
invariance tests a few diagrams were recalculated by hand.

\section{Numerical Results}
\label{sec:results}
Our numerical studies are performed with the set of diquark-model parameters
\begin{equation}
f_{\text{S}} = 73.85 \, \hbox{MeV}, \quad Q_{\text{S}}^2 = 3.22 \,
\hbox{GeV}^2, \quad a_{\text{S}} =
0.15
\; ,
\label{param}
\end{equation}
which has been found by fitting elastic electron-nucleon scattering data
\cite{Ja93} and which
provides also reasonable results in other applications of the diquark model
\cite{Kro93,Kro95}.
A detailed explanation how the convolution integral, Eq.~(\ref{convol}), for
the various n-point
contributions has been treated numerically is given in
Appendix~\ref{sec:proppol}. At this point we
only want to emphasize that propagator singularities have been carefully
separated and integrated
analytically. The remaining integrals could be performed by means of rather
fast fixed-point Gaussian
quadrature.

One of the characteristic qualitative features of perturbative QCD predictions
is the fixed-angle
scaling behaviour of cross sections. Within the diquark model the $\gamma
\text{p} \longrightarrow
{\text{K}^{(\ast)}}^+ \Lambda$ cross section behaves at large $\hat{s}$ like
\widetext
\begin{equation}
\frac{d\sigma}{dt} \propto \hat{s}^{-5} \left[ F_{\text{S}}^{(3)}(- <x_2> <y_2>
\hat{t}) \right]^2
h(\hat{t}/\hat{s}) \stackrel{\hat{s} \rightarrow \infty}{\longrightarrow}
\hat{s}^{-7}
\tilde{h}(\hat{t}/\hat{s})
\, .
\label{scal}
\end{equation}
\narrowtext
\noindent
$<x_2>$ and $<y_2>$ denote average values of the longitudinal momentum fraction
of the diquark in a
proton or $\Lambda$, respectively. Equation~(\ref{scal}) shows that the scaling
behaviour of the pure
quark hard-scattering model \cite{BL89} is recovered in the limit $\hat{s}
\rightarrow \infty$.
However, at finite $\hat{s}$, where the diquark form factor
$F_{\text{S}}^{(3)}$ becomes operational
and diquarks appear as nearly elementary particles, the $\hat{s}^{-7}$
power-law is modified.
Additional deviations from the $\hat{s}^{-7}$ decay of the cross section are
due to logarithmic
corrections (hidden in the functions $F_{\text{S}}^{(3)}$ and $h$) which have
there origin in the
running coupling constant $\alpha_{\text{s}}$ and eventually in the evolution
of the DAs $\phi$
(neglected in our calculation).

\subsection{$\gamma \text{p} \longrightarrow \text{K}^+ \Lambda$}
\label{subsec:K}
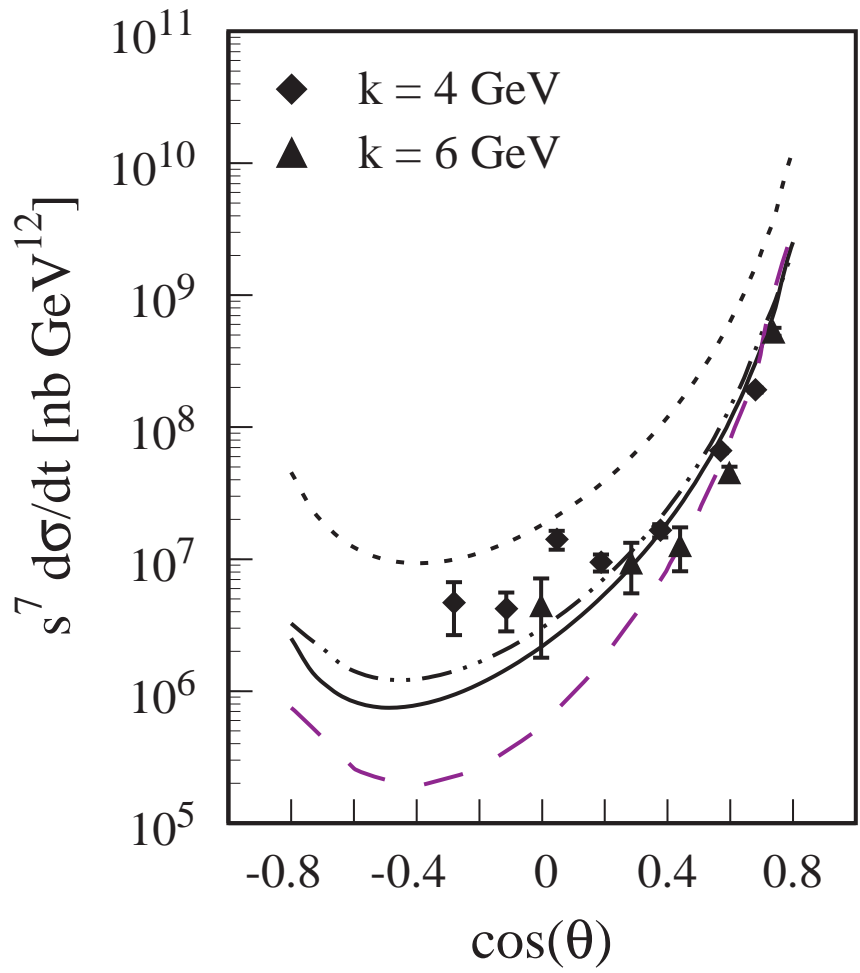
\begin{figure}
\caption{Differential cross section for $\gamma \text{p} \longrightarrow
\text{K}^{+} \Lambda$ scaled by $s^7$ vs. $\cos(\theta_{\text{cm}})$. Solid
(dash-double-dotted) line:
diquark-model result at $p_{\text{lab}}^\gamma = 6 \, \text{GeV}$ $(4 \,
\text{GeV})$, proton
and Lambda DAs chosen according to Eq.\ (\ref{DAp}), Kaon DA according to Eq.\
(\ref{DAKa})
(asymptotic DA); short-dashed line: diquark-model result at
$p_{\text{lab}}^\gamma = 6
\, \text{GeV}$, proton and Lambda DAs chosen according to Eq.\ (\ref{DAp}),
Kaon DA
according to Eq.\ (\ref{DAKCZ})
(Chernyak-Zhitnitsky DA \protect \cite{CZ84}); long-dashed line:
quark-model result \protect \cite{FHZ91} for the asymmetric proton and Lambda
DAs of
Ref.\ \protect \cite{FZOZ88} and the Kaon DA Eq.\ (\ref{DAKCZ}). Experimental
data are taken from
Ref.\ \protect \cite{An76}.}
\label{sigma0K}
\end{figure}

Figure~\ref{sigma0K} shows the diquark-model predictions for $s^7 d\sigma/dt$
along with
the few existing large-momentum transfer data \cite{An76} and the outcome of
the pure quark
hard-scattering model \cite{FHZ91} (long-dashed curve). Whereas the DAs of
proton and $\Lambda$
have been kept fixed according to Eq.~(\ref{DAp}) we have varied the K$^+$ DA.
The solid and
the short-dashed line represent results for the asymptotic (Eq.~(\ref{DAKa}))
and the two-humped
(Eq.~(\ref{DAKCZ})) K$^+$ DA, respectively, evaluated at
$E_{\text{lab}}^{\gamma} = 6$ GeV. The better
performance of the asymptotic DA and the overshooting of the asymmetric DA is
in line with the
conclusion drawn from the investigation of the pion-photon transition form
factor \cite{JKR96,O95} where, for the case of the pion, the CZ DA is
clearly ruled out.
There, strongly end-point concentrated DAs are also overshooting the data.
Our findings have to be contrasted with those obtained
within the pure quark-model calculation of photoproduction
\cite{FHZ91}, where the
asymptotic forms for both, baryon and meson DAs, give systematically larger
results than the
combination of very asymmetric DAs. However, the numerics of Ref.~\cite{FHZ91}
must be taken with
some provisio. For Compton scattering off nucleons it has been demonstrated
\cite{KN91} that the very
crude treatment of propagator singularities adopted in Ref.~\cite{FHZ91},
namely keeping
$i \epsilon$ small but finite, may lead to deviations from the correct result
which are as large as
one order of magnitude. The sensitivity of our calculation to the choice of the
baryon DAs has been
checked only with respect to their end-point behaviour $x \rightarrow 0,1$.
Neglecting the
exponential factor in Eq.~(\ref{DAp}) results in a slight reduction of the
cross section, e.g.
$\approx 8\%$ at $\theta_{\text{cm}} = 90^{\circ}$ and $E_{\text{lab}}^{\gamma}
= 6$ GeV. Deviations from the
scaling behaviour can be estimated by comparing the dash-double-dotted and the
solid curve, which
correspond to the asymptotic K$^+$ DA at $E_{\text{lab}}^{\gamma} = 4$ and $6$
GeV, respectively.

We have also examined the relative importance of various groups of Feynman
graphs and found the
3-point contributions to be by far the most important. 4- and 5-point
contributions amount to
$\approx 5\%$ at $\theta_{\text{cm}} = 90^{\circ}$ and $E_{\text{lab}}^{\gamma}
= 6$ GeV as long as
only $d\sigma/dt$ is considered. Their influence decreases from larger to
smaller angles. Spin
observables, on the other hand, are much more affected by 4- and 5-point
contributions.

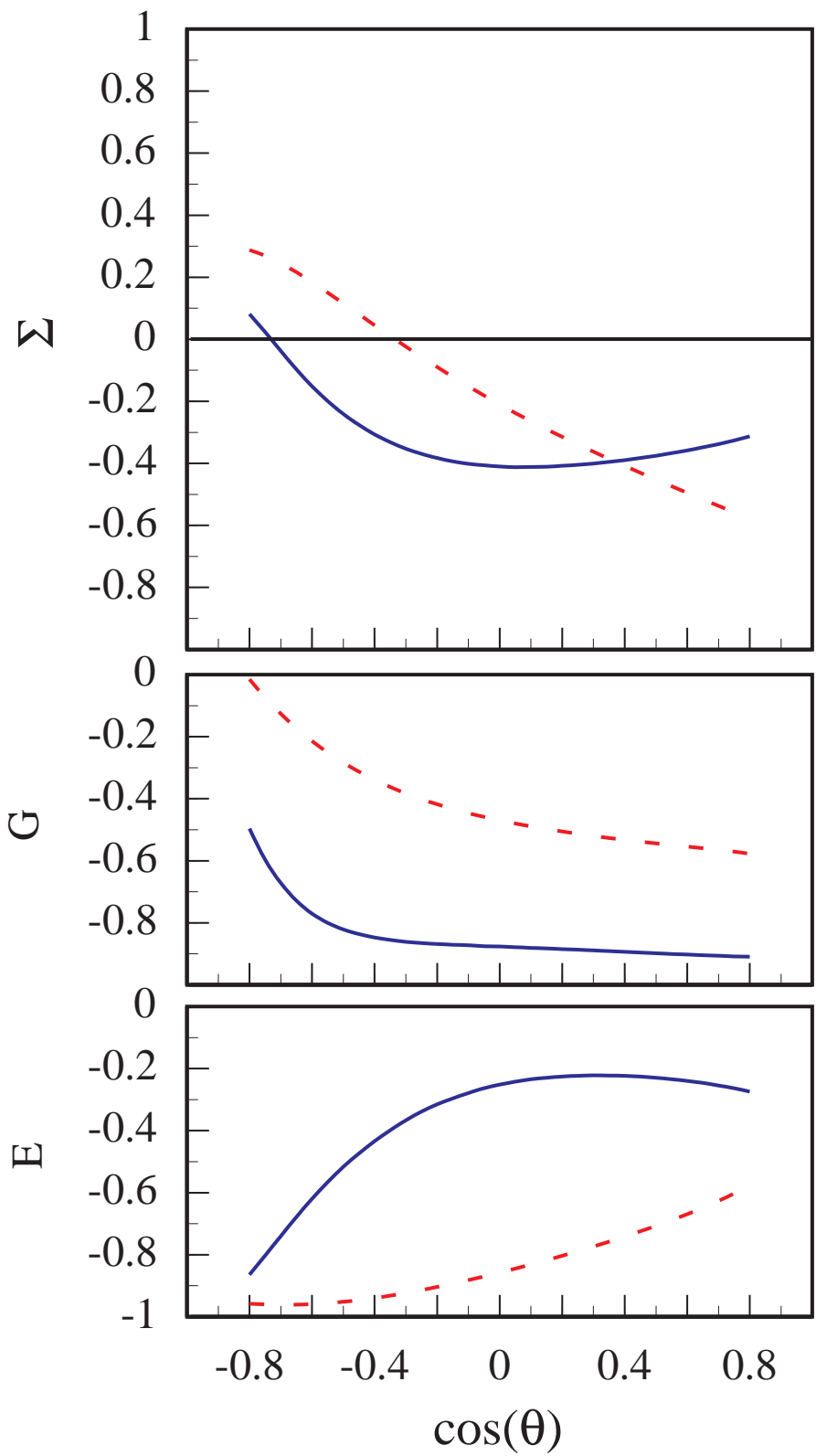
\begin{figure}
\caption{Diquark model predictions for the non-vanishing $\gamma \text{p}
\longrightarrow
\text{K}^{+} \Lambda$ polarization observables. Full (short-dashed) line: same
as in
Fig.~\ref{sigma0K}.}
\label{polobs}
\end{figure}

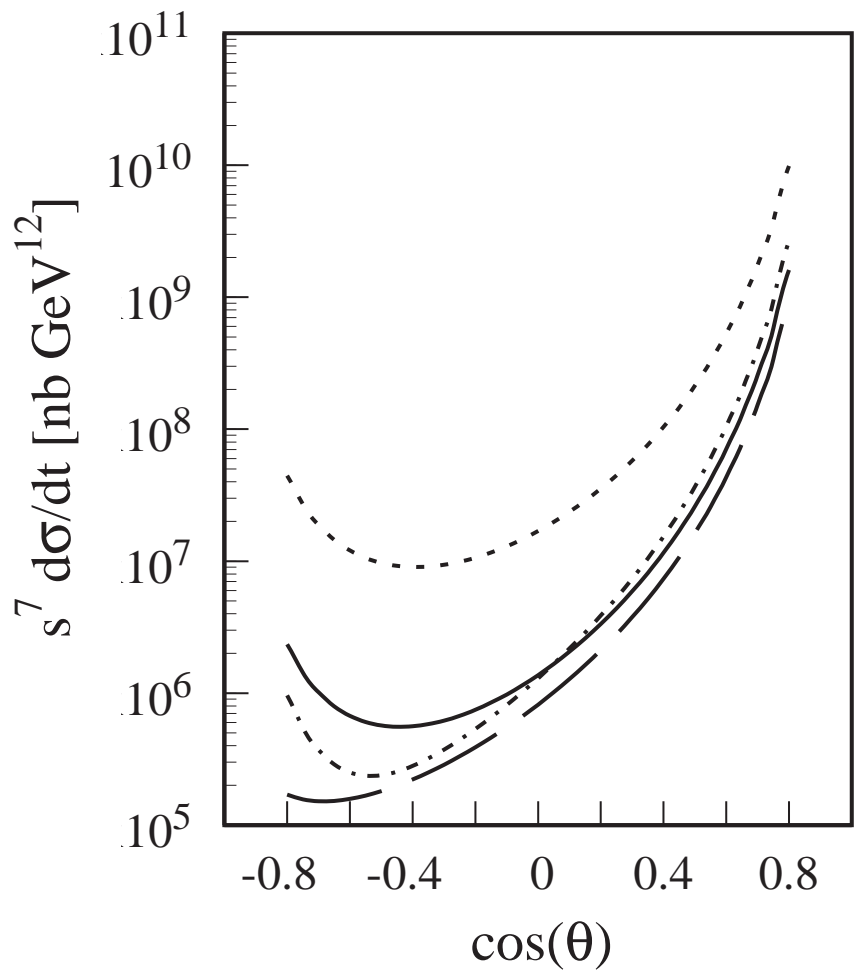
\begin{figure}
\caption{Differential cross section for $\gamma \text{p} \longrightarrow
\text{K}^{+} \Lambda$ scaled by $s^7$ vs. $\cos(\theta_{\text{cm}})$ at
$p_{\text{lab}}^\gamma = 6 \,
\text{GeV}$ -- contributions of the helicity amplitudes S$_1$ and S$_2$,
respectively.
Solid (long-dashed) line: contribution of S$_1$ (S$_2$) for proton and Lambda
DAs chosen
according to Eq.\ (\ref{DAp}), Kaon DA according to Eq.\ (\ref{DAKa});
short-dashed (dash-dotted) line: contribution of S$_1$ (S$_2$) for proton
and Lambda DAs chosen according to
Eq.\ (\ref{DAp}), Kaon DA according to Eq.\ (\ref{DAKCZ}). }
\label{sigmapol}
\end{figure}

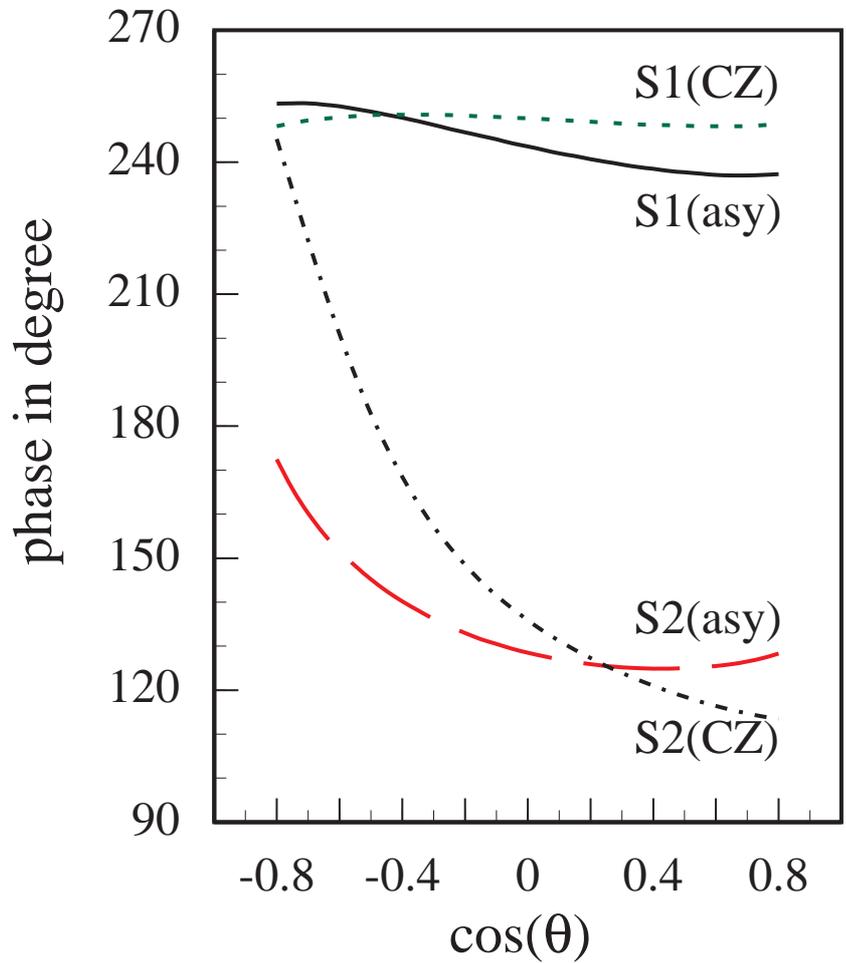
\begin{figure}
\caption{Phases of the helicity amplitudes  S$_1$ and S$_2$, respectively.
Lines as in Fig.\ \ref{sigmapol}.}
\label{phase}
\end{figure}

The three non-vanishing spin observables $E$, $\Sigma$, and $G$ are depicted in
Fig.~\ref{polobs}.
Whereas E measures the relative strength of the two amplitudes $S_1$ and $S_2$,
$\Sigma$ and G
are in addition influenced by the phase difference of these two amplitudes. To
make the interplay of
the two amplitudes $S_1$ and $S_2$ more obvious we have also plotted their
moduli and phases in
Figs.~\ref{sigmapol} and \ref{phase}, respectively. For both choices of the
Kaon DA $S_1$ is observed
to be the dominant amplitude in backward direction. For
$\phi^{\text{K}}_{\text{CZ}}$ it remains
dominant over the whole angular range. In contrast, $S_2$ becomes increasingly
important for
$\phi^{\text{K}}_{\text{asy}}$ if one goes from backward to forward direction.
This behaviour is
clearly reflected by the double-polarization observable $E$. The phase
difference between $S_1$ and
$S_2$ varies rather moderately over the whole angular range for
$\phi^{\text{K}}_{\text{asy}}$,
whereas it changes dramatically for $\phi^{\text{K}}_{\text{CZ}}$.
Unfortunately, the information on
the phase difference is hidden in the photon asymmetry $\Sigma$ and the double
polarization observable
$G$ for which the dependence on the choice of the Kaon DA is not so
aggravating.

Let us recall at this point that the occurrence of nontrivial phases in
photoproduction amplitudes
is a consequence of the fact that most of the Feynman diagrams contain internal
gluons that can
propagate on mass shell in certain kinematic regions of the momentum-fraction
space. The treatment of
the corresponding propagator singularities by means of the usual Feynman
prescription results in an
imaginary contribution to photoproduction amplitudes. One may worry about the
validity of
perturbation theory for a freely propagating gluon which is expected to be
modified by long-distance
effects. But fortunately photoproduction belongs to a class of exclusive
reactions which does, to
leading order perturbative QCD, not require the resummation of gluonic
radiative corrections
(Sudakov effects). As has been proved in Ref.~\cite{FSZ89} the standard
factorization formula,
Eq.~(\ref{convol}), produces already an infrared finite amplitude.

Unlike $\Sigma$, $G$, and $E$, which have not been measured as yet, the
determination of the $\Lambda$
polarization $P$ has been attempted already \cite{V72} (for more recent
efforts, cf.
Ref.~\cite{ELSA95}). The way to determine the transverse polarization of the
$\Lambda$ is to
detect, in addition to the K$^+$, the proton coming from the weak $\Lambda
\longrightarrow
\text{p} \pi^-$ decay. The transverse $\Lambda$ polarization $P$ then follows
from the known ($P$
dependent) angular distribution of the weak $\Lambda \longrightarrow \text{p}
\pi^-$ decay. According
to  the diquark model, and also the pure quark model, $P$ is expected to vanish
in the hard-scattering
regime ($t u / s \gg m_{\text{p}}^2$). The present data, however, are at too
small $s$ and $t$
to allow conclusions about the validity and quality of these perturbative
models. It would be
interesting to see, whether the occurrence of sizable transverse polarizations
at $p_{\perp}
\approx$~few~GeV, as observed e.g. in elastic p-p scattering \cite{C90} or
inclusive production of
hyperons in p-p collisions \cite{B76}, continues to the $\gamma \text{p}
\longrightarrow {\text{K}}^+
\Lambda$ process. This would be an indication that, besides the perturbative
mechanism,
non-perturbative physics (beyond diquarks) is still at work.

We have also computed differential cross sections for the reaction $\gamma
\text{n}
\longrightarrow \text{K}^{0} \Lambda$. For $\cos(\theta_{\text{cm}}) \geq 0$
they are considerably smaller than
the corresponding $\gamma \text{p} \longrightarrow \text{K}^{+} \Lambda$ cross
sections. The amount of
suppression depends on the choice of the K$^0$ DA. For the asymptotic DA the
suppression factor is
$\approx 10$ in the whole forward region, whereas it increases for the
two-humped DA from 2 to
$\approx 10$ when $\cos(\theta_{\text{cm}})$ is varied from $0$ to $0.8$
($E_{\text{lab}}^{\gamma} = 6$ GeV). In
view of the plans at CEBAF to study $\gamma \text{n} \longrightarrow
\text{K}^{0} \Lambda$ by means
of a deuteron target \cite{CEBAF95} this is certainly an interesting
observation which could be
helpful to pin down the uncertainties of the Kaon~DA.

\subsection{$\gamma \text{p} \longrightarrow \text{K}^{\ast +} \Lambda$}
\label{subsec:Ks}
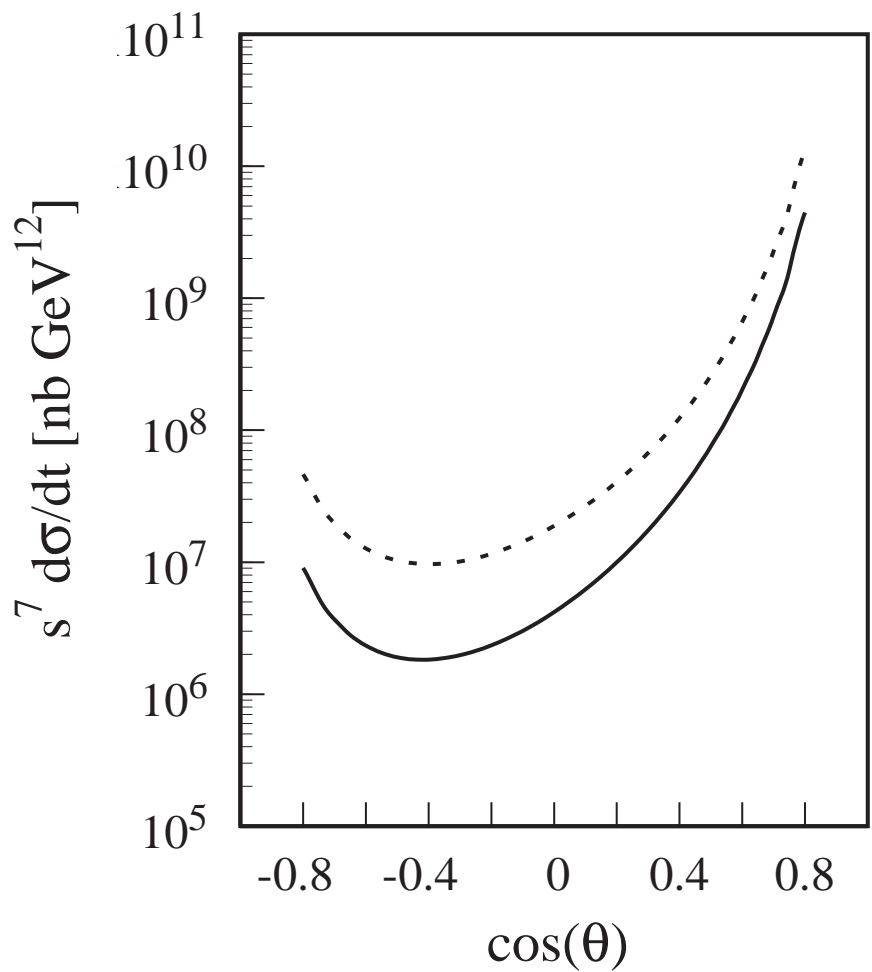
\begin{figure}
\caption{Differential cross section for $\gamma \text{p} \longrightarrow
\text{K}^{\ast +} \Lambda$ scaled by $s^7$ vs. $\cos(\theta_{\text{cm}})$ at
$p_{\text{lab}}^\gamma = 6
\, \text{GeV}$. Solid line: diquark-model prediction, proton and Lambda DAs
chosen according
to Eq.\ (\ref{DAp}), K$^\ast$ DA according to Eq.\ (\ref{DAKa}) (asymptotic
DA); short-dashed
line: diquark-model prediction, proton and Lambda DAs chosen according to Eq.\
(\ref{DAp}),
K$^\ast$ DA according to Eqs.\ (\ref{DAKsCZl}) and (\ref{DAKsCZt}) (taken from
Ref.~\protect
\cite{BC90}).}
\label{sigma0Ks}
\end{figure}

The diquark-model results for $(d\sigma/dt)_{\gamma \text{p} \rightarrow
{\text{K}^{\ast}}^{+}
\Lambda}$ are plotted in Fig.~\ref{sigma0Ks}. Again curves are shown for two
choices of the
${\text{K}^{\ast}}^{+}$ DA with p and $\Lambda$ DA kept fixed
($E_{\text{lab}}^{\gamma} = 6$ GeV).
The results resemble those for photoproduction of K$^+$ mesons. Cross sections
for the asymmetric
DA, Eqs.~(\ref{DAKsCZl}) and (\ref{DAKsCZt}), are nearly one order of magnitude
larger than for the
asymptotic DA, Eq.~(\ref{DAKa}). If $\phi_{\text{asy}}$  is taken for both,
K$^+$ and
${\text{K}^{\ast}}^+$, the photoproduction cross section for the
${\text{K}^{\ast}}^+$ vector
meson is found to be by a factor of 1.8 - 3.6 (depending on the scattering
angle) larger than that
for the pseudoscalar K$^+$ meson. An increase by a factor 1.73 is due to the
different K$^+$ and
${\text{K}^{\ast}}^+$ decay constants $f^{\text{K}}$ and
$f^{\text{K}^{\ast}}$. The remaining
difference is caused by the contribution of transversially polarized
${\text{K}^{\ast}}^+$ mesons,
which increases from small to large scattering angles. The situation would be
quite similar if we had
taken $\phi^{\text{K}}_{\text{CZ}}$ for both, K$^+$ and ${\text{K}^{\ast}}^+$.
However, the
enhancement of the ${\text{K}^{\ast}}^+$ cross section is completely
compensated, if
$\phi^{\text{K}}_{\text{CZ}}$ is replaced by
$\phi^{\text{K}^{\ast}}_{\text{L}}$ and
$\phi^{\text{K}^{\ast}}_{\text{T}}$ (cf. Eqs.~(\ref{DAKsCZl}) and
(\ref{DAKsCZt})) when going from
K$^+$ to ${\text{K}^{\ast}}^+$ photoproduction.

Till now large momentum transfer data for photoproduction of vector mesons are
only available for the  reaction $\gamma \text{p} \longrightarrow (\rho_0
+\omega) \text{p}$
\cite{An76}. At $\theta_{\text{cm}} = 90^{\circ}$ and $E_{\text{lab}}^{\gamma}
\approx 6$ GeV the
experimental value of the cross section ratio $(d\sigma/dt)_{\gamma \text{p}
\rightarrow (\rho_0
+\omega) \text{p}}/(d\sigma/dt)_{\gamma \text{p} \rightarrow \pi_0 \text{p}}$
is $\approx 2$. This
means in particular that $(d\sigma/dt)_{\gamma \text{p} \rightarrow \rho_0
\text{p}} \aleq
2 (d\sigma/dt)_{\gamma \text{p} \rightarrow \pi_0 \text{p}}$. The difference in
the $\rho_0$
and $\pi_0$ decay constants ($f_{\text{decay}}^{\rho} \approx 1.5
f_{\text{decay}}^{\pi}$),
however, already implies an enhancement of the $\rho_0$ photoproduction cross
section as compared
to the $\pi_0$ one by a factor of $\approx 2.3$ which is further magnified by
contributions of
transversially polarized $\rho$s. The only way to compensate part of this
enhancement is to assume
that (as above in the case of K$^+$ and ${\text{K}^{\ast}}^+$) the DAs of $\pi$
and $\rho$ differ from
each other. Experimental data on photoproduction of pseudoscalar and vector
mesons with the same
flavour content could thus be very useful to work out differences in the
corresponding DAs.

\section{Summary and Conclusions}
\label{sec:summary}
We have investigated photoproduction of K-$\Lambda$ and K$^{\ast}$-$\Lambda$
final states in the
few-GeV momentum-transfer region. Our analysis is based on perturbative QCD
supplemented by the
assumption that baryons can be treated as quark-diquark systems. The present
calculation continues
previous work on photon-induced hadronic reactions \cite{Ja93,Kro93,Kro95}
performed within the same
approach. By modeling quark-quark correlations inside a baryon as
quasi-elementary particles -
scalar and vector diquarks -  we account for some non-perturbative effects. In
this way we are able
to extend the range of applicability of the pure quark hard-scattering approach
from large down to
moderately large momentum transfers. The fact that the photoproduction channels
we are interested
in contain a $\Lambda$ in the final state entails a considerable reduction in
computational effort.
In contrast to arbitrary photoproduction reactions only scalar diquarks must be
taken into
consideration. This has the consequence that helicity amplitudes and hence
spin-observables
violating hadronic helicity conservation (cf. Eq.~(\ref{helcondb})), e.g. the
$\Lambda$ polarization, are
predicted to vanish.

Our numerical studies have been performed with the diquark-model parameters and
the quark-diquark DAs
proposed in Ref.~\cite{Ja93}. We have paid special attention to the correct
and numerically robust
treatment of propagator singularities (cf. Appendix~\ref{sec:proppol}).
Reasonable agreement with the
few existing $\gamma \text{p} \longrightarrow \text{K}^+ \Lambda$ data is
achieved already with the
asymptotic form for the K$^+$ DA. On the other hand, the end-point concentrated
K$^+$ DA,
based on QCD sum rules~\cite{CZ84}, seems to perform less well. The
corresponding curve lies far
beyond the data. The difference between these two Kaon DAs is also clearly
visible in the three
non-vanishing polarization observables, i.e. the photon asymmetry $\Sigma$ and
the two
double-polarization  observables G and E. It is most pronounced in the
observable E.
Another quantity which we found to be very sensitive on the choice of the Kaon
DA is the angular
dependence of the cross-section ratio $(d\sigma/dt)_{\gamma \text{n}
\rightarrow \text{K}^0 \Lambda
}/(d\sigma/dt)_{\gamma \text{p} \rightarrow \text{K}^+ \Lambda}$.
For the photoproduction of the K$^{\ast}$ vector meson there are no data to
compare with. We have
again tested the asymptotic DA and a K$^{\ast}$ DA which obeys QCD sum-rule
constraints on the
lowest moments~\cite{BC90}. The differences in the results for the two DAs are
quite similar to those
for photoproduction of the pseudoscalar K$^+$ meson. When going from K to
K$^{\ast}$ photoproduction
the cross section becomes larger due to the different K and K$^{\ast}$ decay
constants and the
additional contributions from transversially polarized K$^{\ast}$s. This
increase of the cross
section, however, is partly compensated if different DAs for K and K$^{\ast}$
are
used.

With regard to future experiments we consider, apart from more
and better large $p_{\perp}$ cross section data, the polarization
measurement of the recoiling $\Lambda$ as one of the most urgent
tasks. A large polarization indicates that the perturbative QCD regime has not
been entered yet. In the perturbative QCD regime the Kaon DAs
could be restricted by means of quantities, like the
photon asymmetry or the cross section ratios $(d\sigma/dt)_{\gamma \text{n}
\rightarrow \text{K}^0
\Lambda}/(d\sigma/dt)_{\gamma \text{p} \rightarrow \text{K}^+ \Lambda}$ and
$(d\sigma/dt)_{\gamma
\text{p} \rightarrow {\text{K}^{\ast}}^+ \Lambda }/(d\sigma/dt)_{\gamma
\text{p} \rightarrow
\text{K}^+ \Lambda}$, which are very sensitive to the choice of the DAs. With a
maximal photon
laboratory energy of (at present) 4~GeV CEBAF \cite{CEBAF95} touches at best
the border of the
hard-scattering domain. More decisive data could be expected from a future
electron facility like
ELFE~\cite{ELFE95} which is designed to explore the
energy range up to 15 GeV
(or even higher) with
a continuous high intensity electron beam.


\appendix

\section{Analytical expressions \protect \\for the hard amplitudes}
\label{sec:eamp}
In this appendix we quote analytical expressions for those hard amplitudes
which describe
the process $\gamma \text{u} \text{S(ud)} \longrightarrow \text{u}
\bar{\text{s}} \text{s}
\text{S(ud)}$ with the $\text{s} \bar{\text{s}}$-pair being in a spin-zero
state. More generally
speaking, these are just the scalar diquark contributions to photoproduction of
pseudoscalar
mesons. According to Eq.~(\ref{Tfg}) the various n-point contributions to these
amplitudes
can be decomposed into gauge invariant functions $f$ and $g$. The functions $f$
and $g$
which determine the hadronic (helicity conserving) amplitude $S_1$ read
\widetext
\begin{eqnarray}
f_{0,-\frac{1}{2},+1,-\frac{1}{2}}^{(3,\text{q})} & = & - C_{\text{F}}^{(1)}
A^{(\text{S},3)}_{\text{T}}
\frac{z_2 \hat{u}}{y_2 z_1 \hat{t}^2} \left[ y_2 \hat{t} + z_2 \hat{u} \right]
\, , \nonumber \\
g_{0,-\frac{1}{2},+1,-\frac{1}{2}}^{(3,\text{q})} & = & - C_{\text{F}}^{(1)}
A^{(\text{S},3)}_{\text{T}}
\frac{\hat{u}}{x_2 y_1 z_1 \hat{s} \hat{t}^2}
\left[ q_4^2 (z_1 \hat{s} + x_2 \hat{t}) - x_2 (x_2 - z_1) \hat{s} \hat{t}
\right] \, ,
\nonumber \\
f_{0,-\frac{1}{2},+1,-\frac{1}{2}}^{(3,\bar{\text{q}})} & = & -
C_{\text{F}}^{(1)} A^{(\text{S},3)}_{\text{T}}
\frac{1}{x_1 x_2 y_1 y_2 z_1 \hat{t}^2 \hat{u}} \left[
y_1 y_2^2 z_2 (x_2 - z_2) \hat{t}^3 \right. \nonumber \\& & +
y_2 ((y_1 - x_1) (y_1 z_2 + y_2 z_1) z_2 + (x_1 y_1 - z_1^2) x_2 y_1 )
\hat{t}^2 \hat{u} \nonumber \\ & & +
((y_1 - x_1) (y_2^2 z_1 - x_2 y_1^2) z_2 + (y_1 z_2 - x_2 z_1) x_2 y_1 y_2 )
\hat{t} \hat{u}^2 \nonumber \\ &
& \left. + x_2 y_1 (y_2 - x_2) z_1 z_2 \hat{u}^3
\right] \, , \nonumber \\
g_{0,-\frac{1}{2},+1,-\frac{1}{2}}^{(3,\bar{\text{q}})} & = & -
C_{\text{F}}^{(1)} A^{(\text{S},3)}_{\text{T}}
\frac{\hat{s}}{x_2 y_1 \hat{t}^2} \left[ (x_2 - z_1) y_2 \hat{t} + y_1 z_1
\hat{u} \right] \,
, \nonumber \\
f_{0,-\frac{1}{2},+1,-\frac{1}{2}}^{(4,\text{q})} & = & - C_{\text{F}}^{(1)}
A^{(\text{S},4)}_{\text{T}}
\frac{\hat{u}}{y_1 z_1 \hat{s} \hat{t}} \left[ x_2 y_2 \hat{t} - D_1^2 \right]
\, , \nonumber \\
g_{0,-\frac{1}{2},+1,-\frac{1}{2}}^{(4,\text{q})} & = & - C_{\text{F}}^{(2)}
A^{(\text{S},4)}_{\text{T}}
\frac{\hat{u}}{x_2 y_1 y_2 z_1 \hat{s} \hat{t}^2}
\left[ y_2 z_2 \hat{s} - x_2 \hat{t} - x_2 z_2 \hat{u} \right] \, , \nonumber
\\
f_{0,-\frac{1}{2},+1,-\frac{1}{2}}^{(4,\bar{\text{q}})} & = &
-C_{\text{F}}^{(1)} A^{(\text{S},4)}_{\text{T}}
\frac{1}{x_1 y_1 z_1 \hat{t} \hat{u}}
\left[ \hat{s} \hat{t} x_2 y_2^2 + D_2^2 (y_2 \hat{t} - y_1 \hat{u})) \right]
\, , \nonumber \\
g_{0,-\frac{1}{2},+1,-\frac{1}{2}}^{(4,\bar{\text{q}})} & = &
C_{\text{F}}^{(2)} A^{(\text{S},4)}_{\text{T}}
\frac{1}{x_1 x_2 y_1 y_2 \hat{t}^2} \left[ g_3^2 (y_2 \hat{s} +
x_2 \hat{u}) + x_1 y_2 \hat{t} \hat{u} \right] \, , \nonumber \\
f_{0,-\frac{1}{2},+1,-\frac{1}{2}}^{(4,\text{S})} & = & -
C_{\text{F}}^{(1)} A^{(\text{S},4)}_{\text{T}}
\frac{y_1 z_2 \hat{s}}{x_1 z_1 \hat{u}}  \nonumber \\
g_{0,-\frac{1}{2},+1,-\frac{1}{2}}^{(4,\text{S})} & = & - C_{\text{F}}^{(1)}
A^{(\text{S},4)}_{\text{T}}
\frac{1}{z_2 \hat{s}} \left[ x_1 \hat{t} + z_2 \hat{s} \right] \, ,
\nonumber \\
f_{0,-\frac{1}{2},+1,-\frac{1}{2}}^{(5,\text{S})} & = & - C_{\text{F}}^{(1)}
A^{(\text{S},5)}_{\text{T}}
\frac{y_1 z_2 \hat{s}}{x_1 z_1 \hat{u}}
\left[ (D_1^2 + x_2 y_1 \hat{t}) (D_2^2 - y_2 \hat{u}) + x_2 y_2 (y_1 -
y_2 + z_1) \hat{t} \hat{u} \right] \, , \nonumber \\
g_{0,-\frac{1}{2},+1,-\frac{1}{2}}^{(5,\text{S})} & = & - C_{\text{F}}^{(2)}
A^{(\text{S},5)}_{\text{T}}
\frac{1}{x_1 y_1 z_1 z_2 \hat{s} \hat{u}}
\left[ x_1 \hat{u} + z_2 (x_1 \hat{u} - y_1 \hat{s})
\right] \, ,
\label{fg1}
\end{eqnarray}
\narrowtext
\noindent
with
\begin{equation}
C_{\text{F}}^{(1)} = \frac{16}{9 \sqrt{3}} \, , \qquad
C_{\text{F}}^{(2)} = \frac{1}{\sqrt{3}} \, ,
\end{equation}
and
\begin{equation}
A^{(\text{S},n)}_{\text{T}} = 128 \pi^2 \sqrt{\pi \alpha} \alpha_{\text{s}}^2
(\hat{t} \hat{u} / \hat{s}) \sqrt{-\hat{t}}
\, F_{\text{S}}^{(n)} ( - x_2 y_2 t) \, ,
\end{equation}
where $\alpha$ denotes the fine-structure constant.

The functions $f$ and $g$ contributing to the hadronic amplitude $S_2$ are
obtained from those
entering $S_1$  by interchange of the Mandelstam variables $\hat{s}
\leftrightarrow \hat{u}$ and
the momentum fractions $x_1 \leftrightarrow y_1$ and $z_1 \leftrightarrow z_2$.
One finds in particular

\vbox{
\begin{eqnarray}
f_{0,+\frac{1}{2},+1,+\frac{1}{2}}^{(3,\text{q})} & = &  -
g_{0,-\frac{1}{2},+1,-\frac{1}{2}}^{(3,\bar{\text{q}})}
(\hat{s} \leftrightarrow \hat{u}, x_1 \leftrightarrow y_1, z_1 \leftrightarrow
z_2) \, , \nonumber \\
g_{0,+\frac{1}{2},+1,+\frac{1}{2}}^{(3,\text{q})} & = & -
f_{0,-\frac{1}{2},+1,-\frac{1}{2}}^{(3,\bar{\text{q}})}
(\hat{s} \leftrightarrow \hat{u}, x_1 \leftrightarrow y_1, z_1 \leftrightarrow
z_2) \, ,
\nonumber \\
f_{0,+\frac{1}{2},+1,+\frac{1}{2}}^{(3,\bar{\text{q}})} & = &  -
g_{0,-\frac{1}{2},+1,-\frac{1}{2}}^{(3,\text{q})}
(\hat{s} \leftrightarrow \hat{u}, x_1 \leftrightarrow y_1, z_1 \leftrightarrow
z_2) \, , \nonumber \\
g_{0,+\frac{1}{2},+1,+\frac{1}{2}}^{(3,\bar{\text{q}})} & = &  -
f_{0,-\frac{1}{2},+1,-\frac{1}{2}}^{(3,\text{q})}
(\hat{s} \leftrightarrow \hat{u}, x_1 \leftrightarrow y_1, z_1 \leftrightarrow
z_2) \, , \nonumber \\
f_{0,+\frac{1}{2},+1,+\frac{1}{2}}^{(4,\text{q})} & = &  -
f_{0,-\frac{1}{2},+1,-\frac{1}{2}}^{(4,\bar{\text{q}})}
(\hat{s} \leftrightarrow \hat{u}, x_1 \leftrightarrow y_1, z_1 \leftrightarrow
z_2) \, , \nonumber \\
g_{0,+\frac{1}{2},+1,+\frac{1}{2}}^{(4,\text{q})} & = &  -
g_{0,-\frac{1}{2},+1,-\frac{1}{2}}^{(4,\bar{\text{q}})}
(\hat{s} \leftrightarrow \hat{u}, x_1 \leftrightarrow y_1, z_1 \leftrightarrow
z_2) \, , \nonumber \\
f_{0,+\frac{1}{2},+1,+\frac{1}{2}}^{(4,\bar{\text{q}})} & = &   -
f_{0,-\frac{1}{2},+1,-\frac{1}{2}}^{(4,\text{q})}
(\hat{s} \leftrightarrow \hat{u}, x_1 \leftrightarrow y_1, z_1 \leftrightarrow
z_2) \, , \nonumber \\
g_{0,+\frac{1}{2},+1,+\frac{1}{2}}^{(4,\bar{\text{q}})} & = &  -
g_{0,-\frac{1}{2},+1,-\frac{1}{2}}^{(4,\bar{\text{q}})}
(\hat{s} \leftrightarrow \hat{u}, x_1 \leftrightarrow y_1, z_1 \leftrightarrow
z_2) \, , \nonumber \\
f_{0,+\frac{1}{2},+1,+\frac{1}{2}}^{(4,\text{S})} & = &  -
g_{0,-\frac{1}{2},+1,-\frac{1}{2}}^{(4,\text{S})}
(\hat{s} \leftrightarrow \hat{u}, x_1 \leftrightarrow y_1, z_1 \leftrightarrow
z_2) \, , \nonumber \\
g_{0,+\frac{1}{2},+1,+\frac{1}{2}}^{(4,\text{S})} & = &  -
f_{0,-\frac{1}{2},+1,-\frac{1}{2}}^{(4,\text{S})}
(\hat{s} \leftrightarrow \hat{u}, x_1 \leftrightarrow y_1, z_1 \leftrightarrow
z_2) \, , \nonumber \\
f_{0,+\frac{1}{2},+1,+\frac{1}{2}}^{(5,\text{S})} & = &  -
f_{0,-\frac{1}{2},+1,-\frac{1}{2}}^{(5,\text{S})}
(\hat{s} \leftrightarrow \hat{u}, x_1 \leftrightarrow y_1, z_1 \leftrightarrow
z_2) \, , \nonumber \\
g_{0,+\frac{1}{2},+1,+\frac{1}{2}}^{(5,\text{S})} & = &  -
g_{0,-\frac{1}{2},+1,-\frac{1}{2}}^{(5,\text{S})}
(\hat{s} \leftrightarrow \hat{u}, x_1 \leftrightarrow y_1, z_1 \leftrightarrow
z_2) \, . \nonumber \\
\label{fg2}
\end{eqnarray}
}

The hard amplitudes for photoproduction of longitudinally polarized
K$^\ast$-mesons are
(up to a sign change in $\hat{T}_{0,-\frac{1}{2},+1,-\frac{1}{2}}$) also
determined by these
expressions. For  photoproduction of transversially polarized K$^\ast$-mesons
the functions $f$ and
$g$  are of similar length and shape and can be obtained from the authors on
request.

\section{Numerical treatment of propagator poles}
\label{sec:proppol}
The numerical difficulties in performing the convolution integral
Eq.~(\ref{convol}) are
mainly caused by the occurrence of propagator singularities in the range of
integration which
give rise to a principal value integral
\begin{equation}
\frac{1}{k^2 + i \epsilon} = \wp \left(\frac{1}{k^2}\right) - i \pi \delta
(k^2) \, .
\label{pv}
\end{equation}
In what follows, the four cases to be distinguished will be discussed
separately.

\subsection{No propagator on shell}
\label{sec:proppol0}
The only contribution to the convolution integral Eq.~(\ref{convol}) exhibiting
no propagator
singularity is the one corresponding to $g_{\{\lambda\}}^{(4,\bar{\text{q}})}$
(cf.
Eq.~(\ref{Tfg})). In these circumstances the convolution integral is easily
performed by means
of 3-dimensional Gaussian quadrature.

\subsection{One propagator on shell}
\label{sec:proppol1}
If only one of the propagators goes on shell within the integration region --
this happens
for $g_{\{\lambda\}}^{(3,\bar{\text{q}})}$, $g_{\{\lambda\}}^{(4,{\text{q}})}$,
$f_{\{\lambda\}}^{(4,\bar{\text{q}})}$, and $g_{\{\lambda\}}^{(5,\text{S})}$ --
 the
corresponding integrals over $x_1$ have the general structure:
\begin{equation}
I^{(k)} (y_1, z_1) = \int_0^1 dx_1 \frac{h(x_1,y_1,z_1)}{k^2 + i\epsilon}
\, . \label{convol1a}
\end{equation}
In order to simplify notations we have neglected helicity labels and the
dependence
on the Mandelstam variables $\hat{s}$ and $\hat{t}$. Furthermore, the
distribution
amplitudes $\phi_{\text{p}}(x_1)$, $\phi_{\Lambda}^{\dagger}(y_1)$, and
$\phi_{\text{K}}^{\dagger}(z_1)$ have been absorbed into the function
$h(x_1,y_1,z_1)$.  The
integral $I^{(k)}$ may be rewritten to give
\begin{eqnarray}
I^{(k)} (y_1, z_1) & = & \int_0^1 dx_1 \frac{h(x_1,y_1,z_1) -
h(x_1^{(k)},y_1,z_1)}{k^2}
\nonumber \\
& + & h(x_1^{(k)},y_1,z_1) \left( \wp \int_0^1 \frac{dx_1}{k^2} - i \pi
\left\vert
\frac{\partial k^2}{\partial x_1} \right\vert^{-1} \right)
\, , \nonumber \\ \label{convol1b}
\end{eqnarray}
where $x_1^{(k)} = x_1^{(k)}(y_1,z_1)$ represents the zero of $k^2$ (considered
as a function
of $x_1$). The first integral in Eq.~(\ref{convol1b}) is now again tractable by
simple
Gaussian quadrature, whereas the principal-value integral can be done
analytically. Since
$I^{(k)}$ is a regular function of $y_1$ and $z_1$ Gaussian integration can
also be applied
to these variables. Analytical expressions for propagator-pole positions,
principal value
integrals and $x_1$-derivatives of propagator denominators are listed in
Table~\ref{tab:singprop}.

\subsection{Two propagators on shell -- propagator poles not coinciding}
\label{sec:proppol2a}
If two propagators $k_1^{-2}$ and $k_2^{-2}$ go on shell one can proceed
similarly as in
the one-pole case, provided the zeroes $x_1^{(k_1)} = x_1^{(k_1)}(y_1,z_1)$ and
$x_1^{(k_2)}
= x_1^{(k_2)}(y_1,z_1)$ of $k_1^2$ and $k_2^2$ (considered as functions of
$x_1$) do not
coincide for fixed $y_1$ and $z_1$, $0 < y_1, z_1 < 1$ arbitrary. This is
guaranteed for the
Feynman diagrams contributing to $f_{\{\lambda\}}^{(3,{\text{q}})}$,
$f_{\{\lambda\}}^{(4,{\text{q}})}$, $f_{\{\lambda\}}^{(4,{\text{S}})}$, and
$g_{\{\lambda\}}^{(4,\text{S})}$. The $x_1$-integrals to be considered have the
general form
\begin{equation}
I^{(k_1,k_2)} (y_1, z_1) = \int_0^1 dx_1 \frac{h(x_1,y_1,z_1)}{(k_1^2 +
i\epsilon)
(k_2^2 + i\epsilon^{\prime})} \, .
\label{convol2a}
\end{equation}
If $x_1^{(k_1)} \neq x_1^{(k_2)}$ a partial fractioning yields
\widetext
\begin{equation}
I^{(k_1,k_2)} (y_1, z_1) = \frac{1}{x_1^{(k_1)}-x_1^{(k_2)}} \left\{
\left( \frac{\partial k_2^2}{\partial x_1} \right)^{-1} I^{(k_1)} (y_1, z_1) -
\left( \frac{\partial k_1^2}{\partial x_1} \right)^{-1} I^{(k_2)} (y_1, z_1)
\right\}
\, ,\label{convol2b}
\end{equation}
\narrowtext
\noindent
i.e., two terms which again can be treated according to Eq.~(\ref{convol1b}).

\subsection{Two propagators on shell -- propagator poles coinciding}
\label{sec:proppol2b}
This is the worst case and shows up in connection with the functions
$g_{\{\lambda\}}^{(3,{\text{q}})}$, $f_{\{\lambda\}}^{(3,\bar{{\text{q}}})}$,
and
$f_{\{\lambda\}}^{(5,\text{S})}$. The general structure of the $x_1$-integrals
is again
that of Eq.~(\ref{convol2a}). However, now it happens that the two propagator
singularities $x_1^{(k_1)}$ and $x_1^{(k_2)}$ (which still depend on $y_1$ and
$z_1$) coincide for a certain value of $y_1$ ($ 0 < y_1 < 1$), $0 < z_1 <1$
fixed. We
denote this value by $y_1^{(k_1,k_2)} = y_1^{(k_1,k_2)}(z_1)$ -- it still
depends on $z_1$.
We note that the partial fractioning can still be performed for arbitrary
values of $y_1$
and $z_1$ as long as $\epsilon$ and $\epsilon^{\prime}$ are kept finite. By
carefully taking
the limit $\epsilon \rightarrow 0$ in the terms containing $k_1^2$ and
$\epsilon^{\prime}
\rightarrow 0$ in the terms containing $k_2^2$ and using Eq.~(\ref{convol1b})
one ends up
with
\widetext

\vbox{
\begin{eqnarray}
I^{(k_1,k_2)} (y_1, z_1)  =  \frac{1}{x_1^{(k_1)}-x_1^{(k_2)} +
i\tilde{\epsilon}}
& \Biggl\{  & \left( \frac{\partial k_2^2}{\partial x_1} \right)^{-1}
\int_0^1 dx_1 \frac{h(x_1,y_1,z_1) - h(x_1^{(k_1)},y_1,z_1)}{k_1^2}  \nonumber
\\
  & + & \left( \frac{\partial k_2^2}{\partial x_1} \right)^{-1}
h(x_1^{(k_1)},y_1,z_1) \left( \wp \int_0^1 \frac{dx_1}{k_1^2} - i \pi
\left\vert
\frac{\partial k_1^2}{\partial x_1} \right\vert^{-1} \right) \nonumber \\
& - & \left( \frac{\partial k_1^2}{\partial x_1} \right)^{-1}
\int_0^1 dx_1 \frac{h(x_1,y_1,z_1) - h(x_1^{(k_2)},y_1,z_1)}{k_2^2} \nonumber
\\
& - & \left( \frac{\partial k_1^2}{\partial x_1} \right)^{-1}
h(x_1^{(k_2)},y_1,z_1) \left( \wp \int_0^1 \frac{dx_1}{k_2^2} - i \pi
\left\vert
\frac{\partial k_2^2}{\partial x_1} \right\vert^{-1} \right)   \Biggr\} \,
.\nonumber \\
\label{convol3a}
\end{eqnarray}
}
In Eq.\ (\ref{convol3a}) $\tilde{\epsilon}$ stands for either $\epsilon$ or
$\epsilon^{\prime}$.
Closer inspection of
$(x_1^{(k_1)}-x_1^{(k_2)})$ (considered as function of $y_1$) reveals that
the zero $y_1^{(k_1,k_2)}$ is quadratic. On the other hand one also finds that
$y_1^{(k_1,k_2)}$ is a single zero of both $h(x_1^{(k_1)},y_1,z_1)$ and
$h(x_1^{(k_2)},y_1,z_1)$. This immediately implies that the real part of
$I^{(k_1,k_2)} (y_1,
z_1)$ is a regular function of $y_1$ and also
$z_1$ so that Gaussian quadrature is again applicable to the corresponding
integrations.
Taking further into account that $({\partial k_1^2}/{\partial x_1})^{-1}$ and
$({\partial k_2^2}/{\partial x_1})^{-1}$ have different signs the imaginary
part of
$I^{(k_1,k_2)} (y_1,z_1)$ can be written as
\begin{eqnarray}
\Im I^{(k_1,k_2)} (y_1,z_1) = & 2 \pi \displaystyle{\left( \frac {\partial^2
(x_1^{(k_1)}-x_1^{(k_2)})} {\partial {y_1}^2} \right)^{-1} \left(
\frac{\partial k_1^2}{\partial x_1}
\right)^{-1}
\left\vert \frac{\partial k_2^2}{\partial x_1} \right\vert^{-1}}
\phantom{MMMMM}
\nonumber \\ & \times  (\tilde{h}(x_1^{(k_1)},y_1,z_1) +
\tilde{h}(x_1^{(k_1)},y_1,z_1)) \displaystyle{\frac{(y_1 - y_1^{(k_1,k_2)})}
{(y_1 - y_1^{(k_1,k_2)})^2 + i \tilde{\epsilon}}} \, ,
\label{convol3b}
\end{eqnarray}
\narrowtext
\noindent
where $h = (y_1 - y_1^{(k_1,k_2)}) \tilde{h}$. Integrating $\Im I^{(k_1,k_2)}$
with respect to
$y_1$ and letting $\tilde{\epsilon} \rightarrow 0$ gives
\begin{equation}
\int_0^1 dy_1 \Im I^{(k_1,k_2)} (y_1,z_1) = \wp \int_0^1 dy_1 \Im I^{(k_1,k_2)}
(y_1,z_1)
\, ,
\label{convol3c}
\end{equation}
i.e., only the principle-value part of the integration survives.
The principle-value integral in Eq.~(\ref{convol3c}) can be treated analogous
to the
principle-value integrals in $x_1$ (cf. Eq.~(\ref{convol1b})).

Proceeding along the steps outlined in this appendix, i.e. carefully separating
the
singular contributions, exploiting delta functions, rewriting principal-value
integrals
as ordinary integrals plus analytically solvable principle-value integrals, it
is
finally possible to do all the numerical integrations by means of fixed-point
Gaussian
quadrature. For our purposes an x-y-z grid of $20 \times 20 \times 24$ turned
out to be
sufficient. Taking instead a $32 \times 32 \times 48$ grid changes the results
by less
than 0.2\%. The numerical calculations were performed on a DEC7000-610 APLPHA
workstation. For the larger grid size the calculation of $d\sigma/dt(\gamma
\text{p}
\rightarrow \text{K}^+ \Lambda)$ took about 1 second per energy point and
angle.

\begin{figure}
\caption{Constituent kinematics for $\gamma \text{p} \longrightarrow
\text{K}^{(\ast)+} \Lambda$. }
\label{kinem}
\end{figure}

\begin{table}
\caption{Propagator-pole positions, $x_1$-derivatives, and principal value
integrals for the
singular propagators occurring in Eq.\ (\ref{Tfg}).
\label{tab:singprop}}
\begin{tabular}{cccc}
$k^2$&$x_1^{(k)}$&$\frac{\partial k^2}{\partial x_1}$&$\wp \int_0^1
\frac{dx_1}{k^2}$\\
\tableline
$g_1^2$ & $\displaystyle\frac{z_1 \hat{t}}{\hat{t} + z_2 \hat{u}}$ & $-
(\hat{t} + z_2 \hat{u})$
&
$ \displaystyle \frac{-1}{\hat{t} + z_2 \hat{u}}\ln \left( \displaystyle
\frac{- z_2 \hat{s}}{ z_1 \hat{t}} \right) $ \\
$g_2^2$ & $\displaystyle \frac{- y_1 \hat{s}}{y_1 \hat{t} + \hat{u}}$ & $ (y_1
\hat{t} + \hat{u})$ &
$ \displaystyle \frac{-1}{y_1 \hat{t} + \hat{u}}\ln \left( \displaystyle
\frac{- y_1 \hat{s}}{ y_2 \hat{u}} \right) $ \\
$q_2^2$ & $\displaystyle \frac{y_2 z_1 \hat{t} + y_1 z_2 \hat{u}}{y_2 \hat{t} +
z_2 \hat{u}}$ & $-(y_2 \hat{t} + z_2 \hat{u})$ &
$ \displaystyle \frac{-1}{y_2 \hat{t} + z_2 \hat{u}}\ln \left( \displaystyle
\frac{- y_2 z_2 \hat{s}}{y_2 z_1 \hat{t} + y_1 z_2 \hat{u}} \right) $ \\
$q_3^2$ & $\displaystyle \frac{y_2 z_2 \hat{t} + y_1 z_1 \hat{u}}{y_2 \hat{t} +
z_1 \hat{u}}$ & $-(y_2 \hat{t} + z_1 \hat{u})$ &
$ \displaystyle \frac{-1}{y_2 \hat{t} + z_1 \hat{u}}\ln \left( \displaystyle
\frac{- y_2 z_1 \hat{s}}{y_2 z_2 \hat{t} + y_1 z_1 \hat{u}} \right) $ \\
$q_4^2$ & $\displaystyle \frac{- y_1 z_2 \hat{s}}{y_1 \hat{t} + z_2 \hat{u}}$ &
$(y_1 \hat{t} + z_2 \hat{u})$ &
$ \displaystyle \frac{-1}{y_1 \hat{t} + z_2 \hat{u}}\ln \left( \displaystyle
\frac{- y_1 z_2 \hat{s}}{y_1 z_1 \hat{t} + y_2 z_2 \hat{u}} \right) $ \\
$q_5^2$ & $\displaystyle \frac{- y_1 z_1 \hat{s}}{y_1 \hat{t} + z_1 \hat{u}}$ &
$(y_1 \hat{t} + z_1 \hat{u})$ &
$ \displaystyle \frac{-1}{y_1 \hat{t} + z_1 \hat{u}}\ln \left( \displaystyle
\frac{- y_1 z_1 \hat{s}}{y_1 z_2 \hat{t} + y_2 z_1 \hat{u}} \right) $ \\
$D_1^2$ & $\displaystyle \frac{y_1 z_1 \hat{t} + y_2 z_2 \hat{u}}{y_1 \hat{t} +
z_2 \hat{u}}$ & $- (y_1 \hat{t} + z_2 \hat{u})$ &
$ \displaystyle \frac{-1}{y_1 \hat{t} + z_2 \hat{u}}\ln \left( \displaystyle
\frac{- y_1 z_2 \hat{s}}{y_1 z_1 \hat{t} + y_2 z_2 \hat{u}} \right) $ \\
$D_2^2$ & $\displaystyle \frac{- y_2 z_1 \hat{s}}{y_2 \hat{t} + z_1 \hat{u}}$ &
$(y_2 \hat{t} + z_1 \hat{u})$ &
$ \displaystyle \frac{-1}{y_2 \hat{t} + z_1 \hat{u}}\ln \left( \displaystyle
\frac{- y_2 z_1 \hat{s}}{y_2 z_2 \hat{t} + y_1 z_1 \hat{u}} \right) $ \\
\end{tabular}
\end{table}

\end{document}